\documentclass[12pt,a4paper]{article}
\usepackage[latin1]{inputenc}
\usepackage{amstext}
\usepackage{latexsym}
\usepackage{amssymb}
\usepackage{amsmath}
\def\R{\mathbb R}

\newcommand{\eps}{\varepsilon}

\newcommand{\fm}{f^-}

\newcommand{\jn}{ j_0}

\newcommand{\txz}{(t, x+z)}
\newcommand{\dz}{\,dz}
\newcommand{\fpn}{ f^+_0}
\newcommand{\fmn}{ f^-_0}

\newcommand{\diffp}{\,dp}
\newcommand{\tdz}{\hat{t}(z)}

\newcommand{\zme}{|z|^{-1}}
\newcommand{\rnp}{\rho_0^+}
\newcommand{\rnm}{\rho^-_0}
\newcommand{\En}{E_0}
\newcommand{\ctme}{(ct)^{-1}}

\newcommand{\xzp}{(x+z, p)}
\newcommand{\dsz}{\,ds(z)}

\newcommand{\cmv}{c^{-4}}

\newcommand{\nxz}{(0, x+z)}
\newcommand{\rn}{\rho_0}

\newcommand{\tdxzp}{(\hat{t}(z), x+z, p)}

\newcommand{\Dd}{D^{[3]}}

\newcommand{\X}{{ X}^\pm}
\renewcommand{\P}{{P}^\pm}

\newcommand{\fon}{f^\circ_0}
\newcommand{\zmd}{|z|^{-3}}

\newcommand{\dst}{\displaystyle}
\allowdisplaybreaks
\begin{document}

\newtheorem{theorem}{Theorem}[section]
\renewcommand{\thetheorem}{\arabic{section}.\arabic{theorem}}  
\newtheorem{definition}[theorem]{Definition}
\newtheorem{deflem}[theorem]{Definition and Lemma}
\newtheorem{lemma}[theorem]{Lemma}
\newtheorem{example}[theorem]{Example}

\newtheorem{remark}[theorem]{Remark}
\newtheorem{remarks}[theorem]{Remarks}
\newtheorem{cor}[theorem]{Corollary}
\newtheorem{pro}[theorem]{Proposition}
\newtheorem{proposition}[theorem]{Proposition}
\newtheorem{assumption}[theorem]{Assumption}

\renewcommand{\theequation}{\thesection.\arabic{equation}}

\title{Post-Newtonian Dynamics at Order 1.5 in the
  Vlasov-Maxwell System} 
\author{{\sc S.~Bauer\footnote{Supported in parts by
        DFG priority research program SPP 1095}}
        \\[2ex] 
        Universit\"at Duisburg-Essen, Fachbereich
        Mathematik, \\ 
        D\,-\,45117 Essen, Germany \\[1ex]
        radiation
        reaction} 
\date{}
\maketitle
\begin{abstract}\noindent
We study the dynamics of many charges interacting
with the Maxwell field. The particles are modeled 
by means of non-negative distribution functions
$f^+$ and $f^-$ representing two species of
charged matter with positive and negative charge,
respectively. If their initial velocities are
small compared to the speed of light, $c$, then in
lowest order, the Newtonian  or classical limit,
their motion is governed by the Vlasov-Poisson 
system. We investigate higher order corrections
with an explicit control on the error terms.
The Darwin order correction, order $|v/c|^2$, has
been proved previously. In this contribution we
obtain the dissipative corrections due to
radiation damping, which are of order $|v/c|^3$
relative to the Newtonian limit. If all particles
have the same charge-to-mass ratio, the
dissipation would vanish at that order.    
\end{abstract}


\setcounter{equation}{0}

\section {Introduction}
In classical electrodynamics it is well known that 
accelerated charges loose energy by radiation and 
there is a large amount of literature concerning 
effective equations which  include effects due to 
radiation damping without giving a completely relativistic 
description of the system of fields and charges.

A similar but more involved situation occurs in the 
theory of general relativity where accelerated 
matter emits gravitational radiation and is thus 
damped.  The probably best studied example is the 
Hulse-Taylor binary pulsar consisting of a strongly 
self-gravitating system of two stars rotating about 
their common center of mass.
Due to the difficulties as the non-linearity and
the necessity of finding appropriate coodinates 
it seems to be out of reach to treat a system like
the one already mentioned within the full
theory. Hence, it is desirable to have effective
equations valid in  certain limits as in the
electromagnetic case. 

In many applications, as e.g. the Hulse-Taylor binary 
pulsar, the occurring velocities are small compared 
to the speed of light. Thus, it is a natural 
strategy to expand the metric in powers of $|v/c|$, 
($c$ denotes the velocity of light).
The contribution in order zero 
corresponds to the non-relativistic limit where
gravity is governed by  Newtonian
theory. Therefore, higher order corrections are
usually addressed as post-Newtonian
approximations. For an overview concerning
post-Newtonian expansions see \cite{bla}. 
Whereas it is relatively straightforward to
establish relations between the full system and
the equations of  approximation, it is much more
difficult  to give the relation between the
solutions of the two sets of equations. While
order zero is done in \cite{ADR} for
asymptotically flat  solutions any further
progress seems to be difficult at this point.

For this reason it seems to be useful to
investigate  the very similar but less involved
system of charged matter coupled to
electromagnetic fields. In \cite{KS1} and
\cite{KS2} the first post-Newtonian approximations
of the Abraham model, a model consisting of single
charged particles coupled to the Maxwell fields
which they create collectively, are considered
yielding the  Darwin corrections, order $|v/c|^2$,
and radiation corrections, order $|v/c|^3$ with
respect to the Newtonian limit. Explicit estimates
of the error terms are given.

In the present paper we choose a model of many
particles governed by a statistical approach. For
sake of simplicity we assume that there are only
two different species of matter  with mass
normalized to unity and charge normalized to plus
unity and minus unity, respectively. These
distributions of the large number of particles in
phase-space are modeled through the non-negative
distribution functions $f^+$ and $\fm$,
$f^\pm=f^\pm(t, x, p)$, depending on time
$t\in\R$, on position $x\in\R^3$ and momentum
$p\in\R^3$. The dynamics is governed by the
relativistic Vlasov-Maxwell system.

\begin{equation}\label{RVMC}
   \left.\begin{array}{lclcrcl}
   \multicolumn{7}{c}{\partial_t
   f^\pm+\hat{p}\cdot\nabla_x  
   f^\pm \pm(E+c^{-1}\hat{p}\times B)\cdot\nabla_p 
   f^\pm=0,} \\[2ex]
   \;c\nabla\times E & = & -\partial_t B, & &
   c\nabla\times B  & 
   = & \partial_t E+4\pi j \\[2ex]
   \quad\, \nabla\cdot E & = & 4\pi\rho, & \quad &
   \nabla\cdot B 
   &  = & 0, \\[0ex]
   \qquad\quad\rho & := & \displaystyle
   \int(f^+-f^-)\,dp, & & j & := & \displaystyle 
   \int\hat{p} (f^+-f^-)\,dp
   \end{array}\right\}\tag{RVM${}_{c}$}
\end{equation}
Here
\begin{equation}\label{hatv-def}
   \hat{p}=(1+c^{-2} p^2)^{-1/2}p\in\R^3
\end{equation}
is the relativistic velocity associated to
$p$. The Lorentz force $E+c^{-1}\hat{p}\times B$
realizes the coupling of the Maxwell fields $E(t,
x)\in\R^3$ and $B(t, x)\in\R^3$ to the Vlasov
equation, and conversely the density functions
$f^\pm$ enter the field equations via the scalar
charge density $\rho(t, x)$ and the current
density $j(t, x)\in\R^3$, which act as source
terms for the Maxwell equations.
The parameter $c$ gives the speed of light for
given units of time and space of the physical
system represented. As usual we shall deal with
the limit of small velocities by letting $c\to
\infty$. Some background on this procedure is
given Section \ref{PN-sec}. In order to give
the Cauchy problem of (\ref{RVMC}) one has to
prescribe initial data for the densities and the
fields at a certain time $t$,  say $t=0$, 
\begin{equation}
  \label{data-def}
  f^\pm(0, x, p)=f^{\circ, \pm}_c(x, p), \quad
  E(0, x)=E^\circ_c(x), \quad B(0,
  x)=B^\circ_c(x). 
\end{equation}
Henceforward in our notation we will only express
the initial data's dependency upon the  light
velocity  by the subscript $c$ while the
dependency of the solution will be suppressed. 

In the next section we shall describe the
post-Newtonian expansions used in this
paper. First we define a ``naive'' post-Newtonian
expansion of (\ref{RVMC}), see
(\ref{exp-def})-(\ref{fields-def}); whereas this
expansion is well defined for all orders in
$c^{-1}$ and the solutions of the expansion
equations up to order $ c^{-2}$ are good
approximations of the solution of the Cauchy
problem of (\ref{RVMC}) if the initial data are
well  adapted, see \cite{schaeffer:86} and
\cite{degond} for the Newtonian limit and
\cite{bauku} and \cite{degondravi} for the
corrections in order $c^{-2}$, it does not include
damping effects due to radiation which occur in
the order $c^{-3}$, see \cite{BKRR}. Hence, we
introduce a more sophisticated expansion up to
order $c^{-3}$ containing a radiation reaction term,
see (\ref{dVPc}) below. Assuming only one species
of matter, say $f^-=0$, this term would vanish
reflecting that there is no radiation in this order
of $c^{-1}$ in that case.  As in the case of single
charges one has to circumvent the occurence of
so-called run-away solutions in the resulting
dynamics. Details are explained in Section
\ref{cm-sec}. We show that the resulting
effective dynamics is well defined, at least
locally in time, and we give some nice further
properties of the solutions, Proposition
\ref{f0-th} and Lemma \ref{f2-le}. Furthermore, we
give an explicit expression of the fields as
function of the  values of the  densities; in this
expansion the fields are not a degree of freedom
anymore but are enslaved by the densities, see
formulas (\ref{radE-rep}) and (\ref{radB-rep}). 

In Section \ref{comp-sec} we state our main
results concerning the comparison of solutions of
our expansion and solutions of (\ref{RVMC}).

The phase-space of the comparison dynamics defined 
in Section \ref{PN-sec} consists of a pair
$f^{\circ, \pm}_c$,  of smooth non-negative
functions with compact support defined on
$\R^3\times \R^3$. The field quantities are to be
computed from this densities e.g. by means of the
formulas (\ref{radE-rep}) and (\ref{radB-rep}). On
the other hand treating the  Cauchy problem of
(\ref{RVMC}) we also have to specify the initial
fields $E^\circ_c$ and $B^\circ_c $ and it is the
question for which choice of initial data the
dynamics of Section \ref{PN-sec} is a good
comparison dynamics and for which not. Surely, it
is possible to choose initial data for the fields
such that the densities of the two dynamics evolve
in completely different ways. In Section
\ref{comp-ad-sec} we will adapt the initial fields for
the Maxwell dynamics from the comparison dynamics,
see formula (\ref{IC}). That means for given data
for the particle densities we compute  special
fields by means of the formulas (\ref{radE-rep})
and (\ref{radB-rep}) and impose these fields upon
the initial values of the Maxwell fields. That has
the advantage that, from a mathematical point of
view, existence and  uniqueness theorems of local
in times solutions for both dynamics are at hand,
see \cite{glstr}, \cite{schaeffer:86},
\cite{bau1}.  We prove that the error between the
solutions of the two systems is of order ${\cal
  O}(c^{-4})$, see Theorem \ref{Hauptsatz}. We
want to mention that in \cite{schaeffer:86},
\cite{degond} and \cite{bauku} the fields are
adapted up to the relevant orders in the same way.   

There are two drawbacks of this method.
Post-Newtonian expansion is in essence an
expansion of the relativistic velocity $\hat{p}$ and
the retardet time $t-c^{-1}|x-y|$. It is clear that
assuming localized sources the expansion of the
retardet time is only a good approximation in the
near zone of the source where $|x-y|\ll c$. This is
reflected in the  fact that the estimates of the
fields in Theorem \ref{Hauptsatz} and
\ref{Hauptsatz2} are only local in the space
variable $x$. For this reason also the adapted
initial fields are only reliable in the near zone,
in particular they are not of finite energy. 
From a more physical point of view it also seem to
be questionable to use the Cauchy problem at
all. Recalling the motivation of post-Newtonian
expansions one is more interested in  localized
systems, isolated from the rest of the world,
which have already evolved for a long time with
small velocities. Therefore the Cauchy problem
might not be the right formulation since it is not
clear how to incorporate these properties into the
initial fields. In physics textbooks isolated
systems are characterized by the absence of
incoming radiation, that is energy coming into the
system from past null infinity by means of
electromagnetic fields, for a rigorous definition
see \cite{cal1}. For given sources fields
free of incoming radiation are usually calculated
by means of the retardet potential. Past
null infinity is that region of space-time which
is reached in the direction of backward
lightcones. In Section \ref{retRVMC-sec} we
consider a familiy of solutions, parametrized by
$c$ of (\ref{RVMC}), passing through $f^{\circ,
  \pm}_c$ at time $t=0$ where in contrast to the
Cauchy problem of (\ref{RVMC}) the electromagnetic
fields are computed by means of the retardet
potentials alone, see (\ref{retRVMC}). Because it
is our goal to model slow systems we assume that
the momenta are bounded uniformly in $c\ge 1$ and
time $t\in\R$, see Assumption
\ref{retRVMC-ass}(b). It is not the aim of this
paper to investigate existence of such solutions
with mathematical rigor, instead  we will just
assume their existence and some nice properties
used in the sequel, see Assumption
\ref{retRVMC-ass}. Note however that in
\cite{cal1} the existence of global solutions is
proved for small $f^{\circ, \pm}_c$. Furhermore it
is shown that such a solution is  also unique in a
certain class of ``nearly  free streaming
solutions'' and is free of incoming
radiation. We also want to emphasize that the
Larmor formula has been proved for this system,
see (\ref{lam-for}) and \cite{BKRR}.
The underlying physical picture is that in the
absence of incoming radiation from outside the
system any solution of (\ref{RVMC}) will approach 
a solution of (\ref{retRVMC}). That means that
solutions of (\ref{retRVMC}) constitute a kind of
initial layer. 

We prove that the error between solutions of our
comparsion dynamics and solutions of
(\ref{retRVMC}) is of order $\cmv$, see Theorem
\ref{Hauptsatz2}. 

In Section \ref{not-sec} we collect some more
notations used in the proofs of Theorem
\ref{Hauptsatz} and Theorem \ref{Hauptsatz2}. The
proof of Theorem \ref{Hauptsatz} is elaborated in
Section \ref{HS-bew} while the somehow cumbersome
computations of some representation formulas for
the fields is outsourced to the Appendix
\ref{append}. The proof of Theorem
\ref{Hauptsatz2} is presented in Section
\ref{main2-th}.

\section{Post-Newtonian expansion}\label{PN-sec}
We adopt the definition  of a post-Newtonian
approximation from \cite{KR2}, see also
\cite{ADR2} for the Einstein case. Therefore
matter and  fields are  described by functions
$(f^\pm(c), E(c), B(c))$ depending on a parameter
$c\in [c_0, \infty)$ giving a one-parameter family
of solutions of (\ref{RVMC}). This means that
$(f^\pm(c), E(c), B(c))$ describes a one-parameter
family of solutions of physical systems which are
represented in parameter-dependent units where the
numerical value of the speed of light is given by
$c$. A more conventional  physical description of
the post-Newtonian expansion would say that in a
fixed system of units the occurring velocities are
small compared to the speed of light. To be more
precise, note that $(f^\pm, E, B)$ is a solution
of (\ref{RVMC}) with $c=\eps^{-1/2}$ if and only if
\begin{eqnarray}
   f^{\pm, \eps}(t, x, p) & = & \eps^{3/2} f^\pm(
   \eps^{3/2} t, \eps x, \eps^{-1/2} p),
   \nonumber \\ 
   E^\eps(t, x) & = & \eps^2 E(\eps^{3/2} t, \eps
   x ), 
   \label{sca-def} \\ 
   B^\eps(t, x) & = & \eps^2 B(\eps^{3/2} t, \eps
   x) 
   \nonumber
\end{eqnarray}
is a solution of (\ref{RVMC}) with $c=1$. In this
scaling the masses of the system remain  unchanged
$$\int\!\!\!\int f^{\pm, \eps}\,dp\,dx =
\int\!\!\!\int f^\pm\diffp\,dx $$  
while the average momenta 
$$\bar{p}^{\,\eps}=\int\!\!\!\int p f^{\pm, 
  \eps}\,dp\,dx=\sqrt{\eps}\int\!\!\!\int p
  f^\pm\,dp\,dx=\sqrt{\eps}\bar{p}$$  
are scaled by $\sqrt{\eps}$. By definition of the
rescaled fields these fields are slowly varying in
their space and time variables. Thus, the limit
  $c\to \infty$ corresponds to an adiabatic limit.  

In this work we treat the speed of light $c$ as a
parameter and study the behavior of the system as
$c\to \infty$, but note that all Theorems can also
be formulated in a parameter independent
fashion. In that case the value of $c$ is fixed,
say $c=1$, and the initial data has to be scaled
according to (\ref{sca-def}), see \cite{bauku} for
details.

We start with a formal expansion of all
coefficients occurring in (\ref{RVMC}) in powers
of $c^{-1}$. 
\begin{eqnarray}
   f^\pm & = &
   f^\pm_0+c^{-1}f^\pm_1
   +c^{-2}f^\pm_2+c^{-3}f^\pm_3+\ldots,   
   \nonumber \\
   E & = &
   E_0+c^{-1}E_1+c^{-2}E_2+c^{-3}E_3+\ldots,  
   \nonumber\\
   B & = &
   B_0+c^{-1}B_1+c^{-2}B_2+c^{-3}B_3+\ldots,  
   \label{exp-def}\\
   \rho & = &
   \rho_0+c^{-1}\rho_1+c^{-2}\rho_2
   +c^{-3}\rho_3+\ldots,  
   \nonumber \\
   j & = &
   j_0+c^{-1}j_1+c^{-2}j_2+c^{-3}j_3+\ldots. 
   \nonumber 
\end{eqnarray}
Moreover, also the initial denities are assumed to
be expandable $f^{\circ, \pm}_c= f^{\circ,
\pm}_0+c^{-1} f^{\circ, \pm}_1+...\,$.  While
considering the Cauchy problem the same is
understood for the electromagnetic initial
fields. Finally $\hat{p}=p-(c^{-2}/2)p^2p+\ldots$
by (\ref{hatv-def}), where $p^2=|p|^2$. These
expansions can be substituted into (\ref{RVMC}).
Comparing coefficients at every order gives a
hierarchy of equations for these coefficients. The
equations in order $k$ will be addressed as the
$k/2$-PN equations and the solutions $(f^{\pm,
k/2PN}=\sum_{j=0}^kc^{-j}f^\pm_j, E^{k/2PN}=...,
B^{k/2PN}=...)$ as the $k/2$-PN approximation
contributing to the fact that in the context of
general relativity  post-Newtonian approximations
are usually counted in orders of $ c^{-2}$.
In order zero the  well known Vlasov-Poisson
system of plasma physics appears. 
\begin{equation}\label{VP}
   \left.\begin{array}{rcl}
   \multicolumn{3}{c}{\partial_t
   f^\pm_0+p\cdot\nabla_x f^\pm_0 
   \pm E_0\cdot\nabla_p f^\pm_0=0,} \\[1.5ex] 
   E_0(t, x) &=&-\displaystyle\int
   |z|^{-2}\bar{z}\,\rho_0(t, x+z)\,dz, \\[2ex] 
   \rho_0 & = & \displaystyle\int
   (f^+_0-f^-_0)\,dp, \\[2ex] 
   f^\pm_0(0, x, p) & = & f^{\circ, \pm}_0(x,
   p),\end{array}\right\}\tag{VP} 
\end{equation}
where \[\bar{z}=z\zme.\] Note that the degrees of
freedom of the electromagnetic fields up to this
order are lost, reflecting the fact that the limit
$c\to \infty$ is singular and the  hyperbolic
field equations become elliptic. We recall that in
\cite{schaeffer:86} it has been shown that as
$c\to\infty$ the solutions of (\ref{RVMC})
approach a solution of the Vlasov-Poisson system
with the rate ${\cal O}(c^{-1})$; see
\cite{asanouk,degond} for similar results and
\cite{lee} for the case of two spatial
dimensions. The respective Newtonian limits
of other related systems are derived in \cite{ADR,calee}.

Concerning a general $k$ we assume that the lower 
order coefficients are already computed.
Then the fields  in order $k$ have to solve 
\[ 
  \begin{array}{rclcrcl}
   \nabla\times E_k & = & -\partial_t B_{k-1}, 
   &\quad& \nabla\cdot E_k & = & 4\pi \rho_k, \\   
   \nabla\times B_k & = & \partial_t E_{k-1} + 
   4\pi j_{k-1}, &\quad & \nabla\cdot B_k & = & 0.   
  \end{array}
\]
The corresponding Vlasov equation is given by 
\[ \partial_t f^\pm_k+p\cdot \nabla_x f^\pm_k \pm
E_0\cdot \nabla_p f^\pm_k  
    = \mp E_k\cdot \nabla_p f^\pm_0+R_k \]
where $R_k$ is depending on $f^\pm_j,\;\nabla_x 
f^\pm_j,\; \nabla_p f^\pm_,\; E_j,\,B_j, \;j=0,
\cdots,k-1$.  
If we assume that $f^{\circ, \pm}_k=0$ for all odd
$k$, what we will do henceforward without 
mentioning, and employing the explicit form of
 $R_k$ it is easy to show that we  consistently
can set 
\begin{equation}
  \label{shadow-def}
  f^\pm_{2l+1}=0,\qquad E_{2l+1}=0,\qquad B_{2l}=0  
\end{equation}
for all $l=0, 1,2, \ldots\;$. 
Granted $E_k$ is known, we can easily solve for 
$f^\pm_k$ using characteristics. 
Note that for all orders $k$ the characteristic
flow is given by the vectorfield $(p, \pm
E_0)$. Thus, for $k\ge 1$ $E_k$ only enters the
Vlasov equation via the right hand side.

Using the vector identity $-\nabla\times \nabla
\times +\nabla\nabla\cdot=\Delta$ 
we can rewrite the field equations.
\begin{eqnarray}\label{fields-def}
  E_{2k} & = & 4\pi\Delta^{-1} (\nabla\rho_{2k}
  +\partial_t j_{2k-2}) +\Delta^{-1} (\partial_t^2
  E_{2k-2})  
  \nonumber\\
  & = &  4\pi \sum_{l=0}^k \Delta^{-1-l}
  \Big[\partial_t^{2l}(\nabla \rho_{2k-2l}
  +\partial_t j_{2k-2l-2})\Big]
  \nonumber \\
  B_{2k+1} & = & \Delta^{-1} (\partial_t^2
  B_{2k-2}) -4\pi\Delta^{-1}(\nabla\times j_{2k}) 
  \nonumber \\
  & = & -4\pi\sum_{l=0}^k
  \Delta^{-1-l} \Big[\partial_t^{2l} \nabla\times 
  j_{2k-2l}\Big], 
\end{eqnarray}
where quantities carrying a negative index are set
to zero.
For given $f^\pm_k$, thus $\rho_k$ and $j_k$ are
given, and if we assume that all densities have
compact support we  can solve these equations
using the convolution with the fundamental
solution of $\Delta^{-1-l}$. Of course, without
boundary conditions these solutions are not
unique, and at least for higher orders these
solutions will not vanish at infinity.
 
Nevertheless, if we choose these fields the
coupled equtions are easily solved by a 
fix-point iteration for $E_k$.  Thus, this (naive)
PN approximation scheme is well defined.
  
According to  this scheme $B_1$ is given by
\begin{equation}
  \label{B1-def}
  B_1(t, x) = \int|z|^{-2}\bar{z}\times j_0(t,
  x+z)\,dz 
\end{equation}
where
\begin{equation}
  \label{j0-def}
  j_0=\int p(f^+_0-f^-_0)\,dp.
\end{equation}
The couple $(f_2, E_2)$ is the  solution of 
\begin{equation}\label{LVP}
   \left.\begin{array}{c}
   \partial_t f_2^\pm+p\cdot\nabla_x
   f^\pm_2-\frac{1}{2} p^2\,p\cdot\nabla_x
   f^\pm_0 
   \pm E_0\cdot\nabla_p f^\pm_2\pm(E_2+p\times
   B_1)\cdot\nabla_p f^\pm_0=0, \\[1ex] 
   E_2(t, x) = \displaystyle \frac{1}{2}
   \int\bar{z}\partial_t^2\rho_0\txz\dz
   -\int\zme\partial_t j_0\txz\dz
   -\int|z|^{-2}\bar{z}\rho_2\txz\dz \\[2ex]
   \rho_2 = \displaystyle \int(f^+_2-f^-_2)\,dp
   \\[2ex] 
   f_2(0, x, p)=f^{\circ,\pm}_2(x,
   p).\end{array}\right\}\tag{LVP} 
\end{equation}
In analogy to the particle model the 1-PN
approximation 
\begin{equation}
  \label{1pnapp-def}
  f^{\pm, 1PN}=f^\pm_0+ c^{-2} f^\pm_2,\qquad
  E^{1PN}=E_0+ c^{-2} E_2,  \qquad B^{1PN}=c^{-1} B_1 
\end{equation}
is also called the Darwin approximation. It is
Hamiltonian in the following sense. If the
conserved energy 
\[{\cal E}=\int\!\!\!\int \sqrt{1+ c^{-2} p^2/2}\,
(f^++f^-)\,dp\,dx +\frac{1}{8\pi}\int
(E^2+B^2)\,dx \]
of (\ref{RVMC}) is expanded according to
(\ref{exp-def}) one obtains the Darwin energy
defined by ${\cal E}_D={\cal E}_{D, kin}+{\cal
  E}_{D, pot}$. The kinetic and potential energy
are  given by
\begin{eqnarray*}
  {\cal E}_{D, kin} & = & \int\!\!\!\int
  (p^2/2- c^{-2} p^4/8) (f^+_0+f^-_0)+
  c^{-2}p^2/2(f^+_2+f^-_2)\,dp\,dx  \quad\text{and}\\
  {\cal E}_{D, pot} & = & \frac{1}{8\pi} \int
  E_0^2 +2 c^{-2} E_0\cdot E_2+ c^{-2} B_1^2\,dx
\end{eqnarray*}
respectively. One can check that ${\cal E}_D$ is
conserved along solutions of the 1-PN
approximation. If in the Cauchy problem of
(\ref{RVMC}) we adapt the initial data
(\ref{data-def}) to suit the data of  
the 1PN approximation the solutions are tracked
down with an error of order $c^{-3}$, see
\cite{bauku}. Hence, the naive post-Newtonian
expansion is valid up to  this order.   
\subsection{Radiation damping in the 1.5 PN
  approximation}
Using the naive expansion, according to
(\ref{shadow-def}) and (\ref{fields-def}) we would
simply have to add
\begin{eqnarray}
  \label{B3-def}
  B_3(t, x) & =
  &-4\pi\Big[\Delta^{-1}\big(\nabla\times
  j_2\big)+\Delta^{-2}\big(\partial_t^2\nabla\times
  j_0\big) \Big](t, x) 
  \nonumber \\
  & = & \int\zme \nabla\times j_2(t, x)\,dz+\frac
  1 2 \int|z|\partial_t^2\nabla\times j_0(t,
  x)\,dz 
\end{eqnarray}
together with the factor $c^{-3}$ to the magnetic
field, where
\[ j_2=\int
p(f^+_2-f^-_2)-\frac{p^2}{2}\,p(f^+_0-f^-_0)\,dp.
\] 
Therefore  the relevant energy ${\cal E}_D$ has
not to be changed in comparison to the 1PN order
and we would remain with a Hamiltonian system.
On the other hand it is known that in the full
relativistic system energy is radiated to null
infinity. In \cite[Theorem 1.4]{BKRR} it is shown
that in the limit $c\to \infty$, corresponding to
small velocities, the total amount of radiated
energy is given by  
\begin{equation}\label{lam-for}
  \frac{2}{3}c^{-3}|\ddot{D}|^2, 
\end{equation}
where $D$ is the dipole moment of the Newtonian
limit of the matter defined by 
\begin{equation}
  \label{dipole-def}
  D(t)=\int x\rho_0(t, x)\,dx.
\end{equation}
This theorem gives a mathematical formulation and
a rigorous proof of the Larmor formula in case of
Vlasov matter. Hence, we should  introduce a
radiation reaction force causing this loss of
energy.  As already suggested in \cite{KR1,KR2} we
modify the Vlasov  equation of the Newtonian
distribution by incorporating a small correction
into the force term, 
\begin{equation}
  \label{dVPc}
  \partial_t f_0^\pm+p\cdot\nabla_x f^\pm_0\pm(E_0
  +\frac{2}{3c^3}\dddot{D})\cdot\nabla_p
  f^\pm_0 =0. 
\end{equation} 
The additional term is the generalization of the
radiation reaction force used in particle models,
see  
\cite[formula (16.8)]{jac}. 
In passing we note that for this system the
``energy''  
\begin{equation}
  \label{SEne-def}
  {\cal E}_{S}= \frac{1}{2} \int\!\!\!\int
  p^2(f^+_0+f^-_0)(t, x, p)\,dp\,dx 
  +\frac{1}{8\pi}\int |E_0(t,
  x)|^2\,dx-\frac{2}{3}c^{-3}\dot{D}\cdot\ddot{D} 
\end{equation}
is decreasing, more precisely one obtains 
\begin{equation}
  \label{Ene-dec}
  \frac{d}{dt}{\cal E}_S = -\frac{2}{3c^3}
  |\ddot{D}(t)|^2, 
\end{equation}
the subscript $S$ referring to the name
``Schott''-energy under which  this energy can be
found in the literature. This decreasing of energy
can be attributed to the effect of radiation
damping. If we remain with the positiv definite
energy of the Vlasov-Poisson system  
\begin{equation}
  \label{energyVP-def}
  {\cal E}_{VP}=\frac{1}{2}\int\!\!\!\!\int 
  p^2(f^+_0+f^-_0)\,dp\,dx +\frac{1}{8\pi}\int 
  |E_0(t, x)|^2\,dx  
\end{equation}
evaluated along solutions of (\ref{dVPc}) the
``friction''  has a definite sign only in the time
average 
\begin{equation}
  \nonumber
   \frac{d}{dt}{\cal E}_{VP} = \frac{2}{3} c^{-3}
   \big(\ddot{D}\cdot\dddot{D}
   -|\ddot{D}|^2\big). 
\end{equation}

\subsubsection{``Unphysical'' solutions and the 
``Reduced Radiating Vlasov-Poisson system''}
\label{cm-sec} 
Introducing system (\ref{dVPc})  one immediately
runs into the problem  that an initial datum  has
to be supplied for $\ddot{D}$ (note that $D(0)$
and $\dot{D}(0)$ are already determined by
$f^{\circ\pm}_0$) and   there is no obvious way to
extract this information from the approximation
scheme. This phenomenon is also  known in the
theory of accelerated, and thus radiating, single
charges and leads to the unphysical so-called
run-away solutions.  In \cite{KS2} it has been
observed that in the particle model this problem
has the structure of a singular geometric
perturbation problem, and the ``physical''
dynamics is obtained on a center-like manifold of
the full dynamics. In order to adopt this language
to the model under consideration here we assume
that we are supplied with a (local in time)
classical solution $(f^\pm_0, E_0)$  of
(\ref{dVPc})  and assume that the support of
$f^\pm_0(t, \cdot, \cdot)$ remains compact for all
$t$ in the interval of existence of the
solution. We define the bare  mass of the charges
by 
\[ M=\int\!\!\!\int(f^+_0+f_0^-)(t, x, p)\,dp\,dx. \]
Mass conservation and charge conservation for both
species easily follow  from (\ref{dVPc}) and
integration by parts,
\[ \partial_t M=0, \qquad \partial_t
\rho_0^\pm+\nabla\cdot j_0^\pm=0, \]
where  $j^\pm_0=\int p f^\pm_0\,dp$ and
$\rho_0^\pm=\int f^\pm_0\,dp$. We denote the
additional degrees of freedom by $y:=\ddot{D}$
and compute with  
\begin{equation}
  \label{eta-def}
  \eta:=(2/3)c^{-3} M  
\end{equation}
and exploiting (\ref{dVPc}) in combination with
integration by parts twice 
\begin{align}
  y & =  D^{[2]}+\eta \dddot{D} \nonumber
\intertext{%
where $D^{[2]}$ is defined by}
  \label{D2-def}
  D^{[2]}(t)& =\int\!\!\!\int E_0(t, x)
  (f^+_0+f^-_0)(t, x, p)\,dp\,dx. 
\end{align}
Thus, we may rewrite (\ref{dVPc}) in a form clearly
showing the structure of a singular perturbation
problem. 
\begin{equation}
  \label{SGPP}
  \left.
    \begin{array}{lcr}
    \dot{f^\pm_0} & = & F^\pm(f^\pm_0, y) \\
    \eta\dot{y} & = & G(f^\pm_0, y)
    \end{array}\right\}\tag{SGPP${}_\eta$}
\end{equation}
where $F^\pm$ and $G$ are  defined by 
\begin{align*}
  F^\pm(f^\pm_0, y)&=-p\cdot\nabla_x 
  f^\pm_0\mp(E_0+M^{-1}(y-D^{[2]}))\cdot \nabla_p 
  f^\pm_0\\ 
  G(f^\pm_0, y) & = y-D^{[2]}.
\end{align*}

In contrast to \cite{KS2} we are dealing with a
phase-space  of infinite dimension.  Thus, the
proof of the existence of invariant manifolds is
hard. We shall return to  that question in a
forthcoming paper. For the moment we shall take
the existence of a smooth invariant manifold for
granted and assume that it is given by means of a
smooth function $h_\eta=h_\eta(f^{\circ}_0)$,
acting on $C^\infty_0(\R^3\times\R^3)\times
C^\infty_0(\R^3\times\R^3)$ and  taking values in
$\R^3$. In this subsection  $f^\circ_0$ denotes
the couple $(f^{\circ, +}_0, f^{\circ,-}_0)$. The
same convention will be used for $f_0=(f^+_0,
f^-_0)$ and $F=(F^+, F^-)$. The manifold ${\cal
  M_\eta}=\{(f^{\circ}_0, h_\eta(f^{\circ}_0)\}$
is invariant under the flow of (\ref{SGPP}) if the
solution of (\ref{SGPP}) subject to the initial
conditions $(f^{\circ}_0,y(0))= (f^{\circ}_0,
h_\eta(f^{\circ}_0))$ satisfies 
\[ y(t)=h_\eta(f_0(t, \cdot, \cdot)) \]
for all times the solution exits. 

We want to establish a dynamics of Vlasov-Poisson
type which is a good approximation of the dynamics
on the manifold. For this reason we assume that we
can expand $h_\eta$ in $\eta$ about 0,
$h_\eta=h_0+\eta h_1+{\cal O}(\eta^2)$. Inserting
$\eta=0$ we have $G(f_0, h_0(f_0, 0))=0$ which
implies   
\begin{equation}
  \label{h0-rep}
 h_0=D^{[2]}. 
\end{equation}
Exploiting this information we find by a formal
calculation  
\begin{align*}
  \eta\dot{y} & = G(f_0, h_0(f_0)+\eta
              h_1(f_0)+{\cal O}(\eta^2))\\ 
              & = \eta \partial_y G(f_0,
              h_0(f_0))h_1(f_0)+{\cal O}(\eta^2) 
\intertext{and on the other hand}
   \eta\dot{y} & =\eta <h_0^\prime(f_0),
              \dot{f_0}>+{\cal O}(\eta^2) 
\end{align*}
which yields
\[ h_1(f_0)=\Big[\partial_y G(f_0,
h_0(f_0))\Big]^{-1}<h_0^\prime, F(f_0,
h_0(f_0))>.\] 
Here ${}^\prime$ denotes the Frech\'et derivative
and $<\cdot, \cdot>$ is the duality pairing,  
\[ <h_0^\prime(f_0), F(f_0, h_0(f_0))> = 
h_0^\prime(f_0)\cdot\int\!\!\!\int
\Big(-p\cdot\nabla_x(\fpn-\fmn)-E_0\cdot
\nabla_p(\fpn+\fmn)\Big)(\cdot, 
x, p)\,dp\,dx.\]
In view of the definition of $G$ and
(\ref{h0-rep}) we have 
\[ y = D^{[2]}+\eta <{D^{[2]}}^\prime,
F(f_0,D^{[2]})>+{\cal O}(\eta^2).\] 
After some straightforward computations we find 
\[ <{D^{[2]}}^\prime, F(f_0,D^{[2]})>=\Dd \]
where
\begin{equation}
  \label{D3-def}
   D^{[3]}(t)=2\int H^+(t, x)j_0^-(t, x)-H^-(t,
   x)j^+_0(t, x)\,dx 
\end{equation}
and 
\begin{equation}
  \label{Fpm-def}
  H^\pm(t, x):=\oint \zmd\big(-3\bar{z}\otimes\bar{z}+{\rm
  id}\big)\rho_0^\pm(t, x+z)\,dz \in  \R^{3\times
  3}. 
\end{equation}

Note that 
\begin{equation}
  \label{H-def}
  H(z)=-3\bar{z}\otimes\bar{z}+{\rm id}
\end{equation}
is bounded on $\R^3\setminus\{0\}$ homogeneous of
degree zero and satisfies
$\int_{|z|=1}H(z)\dsz=0$.
 
Alternatively $\Dd$ may be found by means of the
following formal calculations, using (\ref{dVPc}).
\begin{eqnarray}
  \nonumber
  \dddot{D}(t) &=& \partial_t^2\int j_0(t,
  x)\,dx+{\cal O}(c^{-3}) 
  \\ \nonumber 
  & = & \dot{D^{[2]}}+{\cal O}(c^{-3}) 
  \\ \nonumber
  & = & D^{[3]}(t)+{\cal O}(c^{-3}). 
\end{eqnarray}
Whereas this second derivation is more simple the
first derivation reveals the import connection to
the dynamics on the manifold. 
 
We introduce the ``reduced radiating
Vlasov-Poisson system'' 
\begin{equation}\label{rrVP}
   \left.\begin{array}{c}\partial_t
   f^\pm_0+p\cdot\nabla_x f^\pm_0 
   \pm (E_0+2/3 c^{-3} D^{[3]})\cdot\nabla_p 
   f^\pm_0=0, \\[1ex] 
   E_0(t, x)=-\displaystyle\int
   |z|^{-2}\bar{z}\,\rho_0(t, x+z)\,dz, \\[2ex]
   \rho_0 = \displaystyle \int (f^+-f^-_0)\,dp
   \\[2ex] 
   f^\pm_0(0, x, p)=f^{\circ, \pm}_0(x,
   p),\end{array}\right\}\tag{rrVP${}_c$} 
\end{equation}
where $D^{[3]}$ is defined according to
(\ref{D3-def}) and (\ref{Fpm-def}).

Our first proposition addresses the existence and
uniqueness of local classical solutions of
(\ref{rrVP}). Furthermore, it provides us with some
estimates useful  in the sequel of this paper. In
every (even)  order $k$  of $c^{-1}$ we shall only
consider smooth compactly supported initial data,  
\begin{gather}
  \label{dat-def}
  f^{\circ, \pm}_k\in C^\infty_0(\R^3\times \R^3),
  \quad f^{\circ, \pm}_k\ge 0 
  \\ \nonumber
   f^{\circ, \pm}_k(x,
  p)=0\quad\text{for}\quad|x|\ge
  r_0\quad\text{or}\quad  
  |p|\ge r_0\quad||f^{\circ\pm}_k||_{W^{4,
  \infty}}\le S_0 
\end{gather}
with some $r_0>0$ and $S_0$ fixed. We call the
constants $r_0$ and $S_0$ 'basic' as all bounds
related to $(f^\pm_0, E_0)$ occurring in the
sequel will only depend on $r_0$ and $S_0$. The
following proposition is proved in \cite{bau1}.  
\begin{proposition}\label{f0-th}
  If $f^{\circ,\pm}_0$ satisfies (\ref{dat-def})
  then there exists a constant  $0< \tilde{T}\le
  \infty$ such that the following holds for $c\ge
  1$. 
\begin{enumerate}
\item[(a)]\label{locex}
   There is a unique classical, i.e. $C^1$
   solution $(f^\pm_0, E_0)$ of (\ref{rrVP}) existing
   on a time interval $[0, T_c)$ with
   $\tilde{T}\le T_c\le\infty$.  
\item[(b)]\label{sup-th} 
  For every $T<\tilde{T}$ there is a constant 
  $M_1(T)>0$ such that for all $0\le t\le T$ 
  $$f^\pm_0(t,x, p)=0\quad\text{if}\quad|x|
  \ge M_1(T)\quad{or}\quad|p|\ge M_1(T).$$
\item[(c)]\label{est-th} 
  In fact, $f^\pm_0$ is $C^\infty$ and for every
  $T<\tilde{T}$ there is constant $M_2(T)$, such
  that for all $0\le t\le T$ 
  \[|\partial^\alpha f^\pm_0(t, x, p)|
  +|\partial_t^\beta E_0(t,
  x)|+|\partial_t^\gamma\Dd(t)| \le M_2(T) \] 
  for all  $x\in\R^3$, $p\in\R^3$, $|\alpha|\le 4$
  and $\beta\le 2$ and $\gamma\le 1$ 
\item[(d)]\label{dtD2D3-th}
   For this solution $(f_0^\pm, E_0)$ we have 
   \[ \dot{D}^{[2]}=D^{[3]},\]
   where $D^{[2]}$ is defined according to 
   (\ref{D2-def}). 
\end{enumerate}
\end{proposition}
Note that the constants $\tilde{T}, M_1(T),
M_2(T)$ appearing in Proposition \ref{f0-th} do
only depend on $r_0$ and $S_0$, in particular they
are independent of $c$. 

As the second moment $\int\!\!\!\int p^2(f^+_0+f^-_0)
\,dp\,dx$ cannot be  bounded a priori by using energy 
conservation it seems difficult to prove  global 
existence of classical solutions of (\ref{rrVP}). 
Note that both methods yielding global existence of 
Vlasov-Poisson type systems essentially rely on such 
an a priori bound, see \cite{pfa} or \cite{schae2} and  
\cite{lipe}.

We shall use solutions of (\ref{rrVP}) instead of 
solutions of (\ref{VP})  in order zero of our 
post-Newtonian approximation  and define  
$B_1, (f^\pm_2, E_2)$ and $B_3$ according to 
(\ref{B1-def}), (\ref{LVP}) and (\ref{B3-def})
respectively. It is important to note  that $(f_0,
E_0)$ is the solution of (\ref{rrVP}). In
particular this solution is depending on $c$ and
thus all other  quantities defined here are
depending on $c$. Concerning the solvability of
(\ref{LVP}) we have the following lemma whose
proof is sketched in \cite{bau1}. 
\begin{lemma}\label{f2-le}
 Let $f^{\circ, \pm}_k$, $k=0, 2$ satisfy
 (\ref{dat-def}). Suppose that $(f^\pm_0, E_0)$ is
 the solution of (\ref{rrVP}). Compute $B_1$
 according to  (\ref{B1-def}). Then (\ref{LVP})
 has a unique  classical solution $(f_2^\pm, E_2)$
 existing on $[0, T_c)$ and enjoying the following
 properties. 
 \begin{enumerate} 
  \item[(a)] \label{sup-le} 
    For every $T<\tilde{T}$ there is a constant  
    $M_3(T)>0$, such that for all $0< t\le T$
    $$ f^\pm_2(t, x, p)=0\quad{if}\quad|x|\ge
    M_3(T) \quad\text{or}\quad|p|\ge M_3(T). $$  
  \item[(b)] \label{est-le} 
    In fact, $f^\pm_2$ is $C^\infty$ and there is
    a constant $M_4(T)$, such that for all $0\le
    t\le T$ 
    $$|\partial^\alpha f^\pm_2(t, x, p)|\le M_4(T)
    \quad\text{for all}\quad x\in\R^3,
    p\in\R^3\quad\text{and}\quad|\alpha|\le 2$$  
\end{enumerate}
\end{lemma}

In the following section we want to show that 
\begin{eqnarray}
  \label{rad-def}
  \nonumber
  f^\pm_R &=& f^\pm_0+c^{-2}f^\pm_2
  \\
  E^R     &=& E_0+c^{-2}E_2+2/3 c^{-3}D^{[3]}
  \\ \nonumber
  B^R     &=& c^{-1}B_1+c^{-3} B_3 
\end{eqnarray}
yields a higher order pointwise approximation of
(\ref{RVMC}) than the Vlasov-Poisson or the Darwin
system defined in \cite{bauku}.  We call
(\ref{rad-def}) the radiation approximation. In
the terminology of post-Newtonian approximations
it is the 1.5PN  approximation. Employing the
Vlasov equation and integration by parts it is not
difficult to prove the following formulas. 
\begin{cor}
  The fields $E^R$ and $B^R$ can be written as
  \begin{subequations}
    \begin{eqnarray}
      \label{radE-rep}
      \nonumber
      E^R(t, x)  & = &   -\int|z|^{-2}\bar{z}(\rho_0+
      c^{-2}\rho_2)(t, x+z)\,dz 
      \\ \nonumber
      & & +
      c^{-2}\frac{1}{2}\int\!\!\!\!\int|z|^{-2}
      \bigg\{3(\bar{z}\cdot   
      p)^2\bar{z}-p^2\bar{z}\bigg\} (f^+_0-f^-_0) 
      (t, x+z, p)\,dp\,dz  
      \\ \nonumber 
      & & -
      c^{-2}\int\!\!\!\!\int\zme
      \bigg\{\bar{z}\otimes\bar{z}+1 
      \bigg\}
      \nonumber \\ 
       & & \hspace{4em}\Big(E_0(t, 
      x+z) +c^{-3}2/3 D^{[3]}(t)\Big) (f^+_0
      +f^-_0)(t, x+z, p)\,dp\,dz 
      \nonumber \\
      & & +c^{-3}\frac{2}{3}\Dd(t)
   \end{eqnarray}
  and 
  \begin{eqnarray}
    \label{radB-rep}
    B^R(t, x)  & = &  c^{-1}\int\!\!\!\!\int
    |z|^{-2} \bar{z}\wedge
    p(f^+_R-f^-_R)(t, x+z, p)\,dp\,dz
    \\ \nonumber
    & & -c^{-3}\frac{3}{2}\int\!\!\!\!\int
    |z|^{-2}(\bar{z}\cdot p)^2\bar{z}\wedge p 
    (f^+_0-f^-_0)(t, x+z, p)\,dp\,dz
    \\ \nonumber 
    & & +\frac{c^{-3}}{2}\int\!\!\!\!\int
    \zme\bigg\{(\bar{z}\wedge p)\otimes 
    \bar{z}+(\bar{z}\cdot p)\bar{z}
    \wedge(\cdots)\bigg\}
    \\ \nonumber 
    & & \hspace{7em}\Big(E_0(t, 
    x+z) +c^{-3}2/3 D^{[3]}(t)\Big) (f^+_0
    +f^-_0)(t, x+z, p)\,dp\,dz 
    \\ 
    & & -c^{-3}\int\bar{z}\wedge \big(H^+(t,
    x+z)j^-_0(t, x+z)-H^-(t, x+z)j^+_0(t,
    x+z)\big)\,dz. 
    \nonumber 
  \end{eqnarray}
  \end{subequations}
\end{cor}
Closing this section we want to mention that there
is another variant of a damped Vlasov Poisson type
system investigated in \cite{KR1} and
\cite{KR2}. While for that system a global
solution theory is at hand, the author did not
succeed in comparing approximations based on
solutions of that system on the one hand and
solutions of the full system on the other hand.  
\section{Comparison of the 1.5 PN dynamics with 
the Vlasov-Maxwell dynamics }\label{comp-sec}
\subsection{The Vlasov-Maxwell dynamics with
  adapted initial values}\label{comp-ad-sec}
For achieving the improved approximation property
we match the initial data of (\ref{RVMC}) 
by the data for radiation system. For prescribed 
initial densities $f^{\circ, \pm}_k$, 
$k=0,2$ we are able to calculate $(f^\pm_0, E_0)$, 
$B_1$,  $(f^\pm_2, E_2)$ and $B_3$ according to 
what has been outlined in Section \ref{PN-sec}. 
We then consider the Cauchy problem of
(\ref{RVMC}) where the initial values are given by
\begin{equation}\label{IC}\tag{IC}
   \left.\begin{array}{lcl}
   f^\pm(0, x, p) & = & f^{\circ, \pm}_c(x, p) 
   = f^{\circ, \pm}_0(x, p)+ c^{-2} f^{\circ, \pm}_2
   (x, p)+c^{-4}f^{\circ, \pm}_{c, free} , \\[1ex]
   E(0, x) & = & E^\circ_c(x):=E_0(0, x)
   +c^{-2}E_2(0, x)+\frac{2}{3}c^{-3}D^{[3]}(0)
   +c^{-4}E^\circ_{c, free} \\[1ex]
   B(0, x) & = & B^\circ_c(x):=c^{-1}B_1(0, x)+
   c^{-3}B_3(0, x)+c^{-4}B^\circ_{c, free} . 
\end{array}\right\}
\end{equation}
In contrast to the contributions in the orders 
0 to 3, which are fixed by the values of the  
approximations,
$(f^{\circ, \pm}_{c, free}, E^\circ_{c, free},
B^\circ_{c, free})$
 are to be chosen freely only subject to the 
constraints $\nabla\cdot E^\circ_{c, free}=4\pi 
\int (f^{\circ, +}_{c, free}-f^{\circ, -}_{c,
  free})\,dp$ and $\nabla\cdot B^\circ_{c,
  free}=0$. Note that the constraint equations in
the lower orders are satisfied by
fiat. Furthermore, we shall assume that the
following bounds hold uniformly in $c$.
\begin{gather}
  \nonumber
  f^{\circ, \pm}_{c, free}\in C^\infty(\R^3\times\R^3),
  \quad E^\circ_{c, free}, B^\circ_{c, free}
  \in C_0^\infty(\R^3)
  \nonumber \\
  f^{\circ, \pm}_{c, free} = 0\quad\text{if}\quad|x|
  \ge r_0\quad\text{or }\quad|p|\ge r_0
  \label{freedat-def}  \\
  ||f^{\circ, \pm}_{c, free}||_{L^\infty} \le  S_0
  \nonumber \\
  ||E^\circ_{c, free}||_{W^{1, \infty}}+
  ||B^\circ_{c, free}||_{W^{1, \infty}} \le S_0.
  \nonumber
\end{gather}
Before we formulate our theorem, let us recall 
that solutions of (\ref{RVMC}) with initial 
data (\ref{IC}) exist at least on some time 
interval $[0, \hat{T})$, which is independent 
of $c\geq 1$; see \cite[Thm.~1]{schaeffer:86}.
\begin{proposition}\label{schaeffer-prop}
Assume that $f^{\circ, \pm}_k, k=0, 2$ satisfies
(\ref{dat-def}). If $f^{\circ, \pm}_c$,
$E^\circ_c$ and $B^\circ_c$ are defined according
to (\ref{IC}), then there exits
$0<\hat{T}\leq\infty$ (independent of $c$) such
that for all $c\geq 1$ the system (\ref{RVMC})
with initial data (\ref{IC}) has a unique
$C^1$-solution $(f, E, B)$ on the time interval
$[0, \hat{T})$. In addition, for every
$0<T<\hat{T}$ there are constants $M_5(T), M_6(T)$
(independent of $c$) such that 
\begin{subequations}
\begin{gather}
   f^\pm(t, x, p) = 0\quad\mbox{if}\quad|x|\ge
   M_5(T) 
   \quad\text{or}\quad |p|\ge M_5(T),
   \label{SchrankeSupport} \\
  |f^\pm(t, x, p)|+ |E(t, x)|+|B(t, x)|\le M_6(T), 
   \label{SchrankeFelder}
\end{gather}
\end{subequations}
for all $x,\,p\in\R^3$, $t\in [0, T]$, and $c\geq
1$. 
\end{proposition}

Actually, in \cite[Thm.~1]{schaeffer:86}
$E^\circ_c$ and $B^\circ_c$ do not depend on $c$
but an inspection of the proof shows that the
assertions remain valid for initial fields defined
by (\ref{IC}). After these preparations we can
state the first of our main results.  
\begin{theorem}\label{Hauptsatz}
Assume that $f^{\circ,\pm}_k, k=0,2$ satisfies
(\ref{dat-def}). From $f^{\circ, \pm}_k$ calculate
$(f_0^\pm, E_0)$, $B_1$, $(f_2^\pm, E_2)$ and
$B_3$, choose initial data $(f^{\circ, \pm}_c,
E^\circ_c, B^\circ_c)$ for (\ref{RVMC}) according
to  (\ref{IC}) and (\ref{freedat-def}). Let $(f,
E, B)$ denote the solution of (\ref{RVMC}) with
initial data (\ref{IC}) and let $(f^\pm_R, E^R,
B^R)$ be defined as in (\ref{rad-def}). Then for
every $T<\min\{\tilde{T}, \hat{T}\}$ and $R>0$
there  are  constants $M(T)>0$ and $M(T, R)>0$,
such that 
\begin{eqnarray}
   |f^\pm(t, x, p)-f^\pm_R(t, x, p)| & \le &
   M(T)c^{-4}\quad\hspace{0.55em} (x\in\R^3),  
   \nonumber \\
   |E(t, x)-E^R(t, x)| & \le & M(T,
   R)\,c^{-4}\quad (|x|\le R), \label{diff-esti}
   \\ 
   |B(t, x)-B^R(t, x)| & \le & M(T,
   R)\,c^{-4}\quad (|x|\le R), \nonumber 
\end{eqnarray}
for all $p\in\R^3$, $t\in [0, T]$, and $c\geq 1$. 
\end{theorem}
The constants $M(T)$ and $M(T, R)$ are independent
of $c\geq 1$, but do depend on the basic constants
$r_0, S_0$.  Note that if (\ref{RVMC}) is compared
to the Vlasov-Poisson system (\ref{VP}) only, one
obtains the estimate  
\begin{align*}
  |f(t, x, p)-f_0(t, x, p)| +|E(t, x)-E_0(t,
  x)|+|B(t, x)| & \le M(T)c^{-1},  
\intertext{see \cite[Thm.~2B]{schaeffer:86}, and
  if compared to the Darwin system the 
estimates} 
  |f(t, x, p)-f^D(t, x, p)|+|B(t, x)-B^D(t, x)|
  &\le M(T)c^{-3}\\ 
  |E(t, x)-E^D(t, x)| &\le M(T, R)c^{-3},  
\end{align*}
 see \cite[Thm.~1.1]{bauku}.
On first glance it could seem that the time
interval $[0, T]\subset [0, \min\{\tilde{T},
\hat{T}\})$, which might be very small, is a
strong limitation of the theorem. But if the
theorem is formulated in an $\eps$-depending
fashion  using fixed units as indicated in
Section \ref{PN-sec} and elaborated in
\cite{bauku} the approximation is valid on the
time interval $[0, \eps^{-3/2}T]$ and thus for
long times on that time scale. 

\subsection{The retardet Vlasov-Maxwell dynamics}
\label{retRVMC-sec}
Following \cite{cal1} we introduce the retardet
relativistic Vlasov-Maxwell system. 
\begin{equation}\label{retRVMC}
  \left. \begin{array}{c}
  \partial_t f^\pm+\hat{p}\cdot\nabla_x f^\pm
  \pm(E+c^{-1}\hat{p}\times B)\cdot\nabla_p
  f^\pm=0,   
  \\[1ex]
  \dst E(t, x) =
  -\int\frac{dy}{|x-y|}\big(\nabla\rho+
  c^{-2}\partial_t j\big)(t-c^{-1}|x-y|, y) 
  \\[1ex]
  \dst B(t, x) =
  c^{-1}\int\frac{dy}{|x-y|}\nabla\times
  j(t-c^{-1}|x-y|, y) 
  \\[1ex] 
  \displaystyle\rho  =  \int (f^+-f^-)\,dp, \qquad
  j =  \int\hat{p} (f^+-f^-)\,dp 
   \end{array}\right\}\tag{${}_{{\rm
  ret}}$RVM${}_{c}$} 
\end{equation}
If we assume that $f^\pm$ is a global $C^1$
solution of (\ref{retRVMC}) and that also $E$ and
$B$ are $C^1$,  then, by means of the Vlasov
equation, $\rho$ and $j$ satisfy the continuity
equation  
\begin{equation}
  \label{cont-eq}
  \partial_t\rho+j=0,  
\end{equation}
and therefore the retardet fields are a solution
of  Maxwell's equations. Thus, $(f, E, B)$ also
solves (\ref{RVMC}). Note that it is necessary to
know the densities for all  time $(-\infty, t]$ in
order to compute the fields at time $t$. Hence,
there is no sense in the notation of a local
solution of this system. As in the case of the
Cauchy problem every solution of (\ref{retRVMC})
satisfies the identity 
\begin{equation}\label{fpmdarst}
   f^\pm(t, x, p) = f^{\pm} (0, X^\pm(0;t,x,p),
   P^\pm(0;t,x,p)),  
\end{equation}
where $s\mapsto (X^\pm(s; t, x, p), P^\pm(s; t, x,
p))$ solves the characteristic system 
\begin{equation}\label{olan}
   \dot x = \hat p,\quad\dot p =
   \pm(E+c^{-1}\hat{p}\times B), 
\end{equation}
with data $X^\pm(t; t, x, p)=x$ and $P^\pm(t; t,
x, p)=p$. For this reason  
\begin{equation}
  \label{finf-est}
0 \leq f^\pm(t,x,p)
\leq {\|f^{\pm}(0, \cdot, \cdot)\|}_\infty.  
\end{equation}
In order to derive our results on
PN-approximations let us assume that we are
furnished with a one-parameter familiy of
solutions (\ref{retRVMC}) satisfying certain
plausible a priori bounds and smoothness
conditions. We consider solutions of
(\ref{retRVMC}) passing through a certain
density configuration at time $t=0$ 
\begin{equation}
  \label{retfini-def}
 f^{\circ, \pm}_c=f^{\circ,\pm}_0+ c^{-2}
 f^{\circ,\pm}_2+ c^{-4} f^{\circ, \pm}_{c, free} 
\end{equation}
where $f^{\circ,\pm}_k, k=0, 2$ and $f^{\circ,
  \pm}_{c, free}$ satisfy (\ref{dat-def})  and
(\ref{freedat-def}) respectively with some
constants $r_0$ and $S_0$. 

\begin{assumption}\label{retRVMC-ass}
\begin{itemize}
\item[(a)] For each $c\ge 1$ there is a solution
  $f^\pm\in C^4(\R\times\R^3\times\R^3)$ of
  (\ref{retRVMC}), passing through
  $f^{\circ, \pm}_c$ at time $t=0$, i.e. 
  $$ f^\pm(0, x, p)=f^{\circ, \pm}_c(x, p), \quad  
  x\in\R^3,\,p\in\R^3.$$
\item[(b)] There exists a constant  $P_1>0$ such 
  that $f^\pm(t, x, p)=0$ for $|p|\ge P_1$ and  
  $c\ge 1$. In particular, $f^\pm(t, x, p)=0$ for 
  $|x|\ge r_0+P_1|t|$ by (\ref{olan}).
\item[(c)] For every $T>0$, $R>0$, and $P>0$ there
  is a  constant  $M_7(T, R, P)>0$ such that
  \[ |\partial_t^{\alpha+1} f^\pm(t, x, p)| 
  +|\partial_t^\alpha\nabla_x f^\pm(t, x, p)|\le 
  M_7(T, R, P) \]
  for $|t|\le T$, $|x|\le R$, $|p|\le P$, and
  $\alpha=0,...,3$, uniformly in $c\geq 1$.
\end{itemize}
\end{assumption}  
After these preparations we can state our second
main result.
\begin{theorem}[Approximation of the retardet
  Vlasov-Maxwell system]\label{Hauptsatz2}
  \hfill{}   

Assume that $f^{\circ,\pm}_k, k=0,2$ satisfies
(\ref{dat-def}). From $f^{\circ, \pm}_k$ calculate
$(f_0^\pm, E_0)$, $B_1$, $(f_2^\pm, E_2)$, $B_3$
and $\Dd$. Define $(f^\pm_R, E^R, B^R)$ according
to (\ref{rad-def}). Assume that $(f^\pm, E, B)$ is
a family of solutions of (\ref{retRVMC})
satisfying Assumption \ref{retRVMC-ass}  with
constants $P_1$ and $M_7(T, R, P)$. Take
$\Tilde{T}$ from Proposition \ref{f0-th}. Then for
every $T<\tilde{T}$ and $R>0$ there are constants
$M(T)$ and $M(T, R)$ such that  
\begin{eqnarray}
   |f^\pm(t, x, p)-f^\pm_R(t, x, p)| & \le &  M(T)
   c^{-4}\quad\hspace{0.55em} (x\in\R^3), 
   \nonumber \\
   |E(t, x)-E^R(t, x)| & \le & M(T, R)\,c^{-4}
   \quad  (|x|\le R), \label{diff-esti2} \\
   |B(t, x)-B^R(t, x)| & \le & M(T, R)\,c^{-4}
   \quad  (|x|\le R), \nonumber
\end{eqnarray}
for all $p\in\R^3$, $t\in [0, T]$, and $c\geq
2P_1$.

The constants $M(T)$ and $M(T, R)$ do only depend
on $r_0, S_0, P_1$ and $M_7(\cdot , \cdot,
\cdot)$. In par\-ticular they are independent of
$c\ge P_1$. 
\end{theorem}

\subsection{Comparison with the particle model}
We shall compare our results for the Vlasov  model
with the results for the particle model governed
by  the Abraham system obtained in
\cite{KS1}  and \cite{KS2}. Both systems share the
features of Hamiltonian approximations up to 1-PN
order and dissipative corrections in the 1.5-PN
approximation leading to an increase of the
phase-space. The right comparison dynamics is
given  on a center-like manifold, where the
dynamics  on this manifold can be approximated
by a modified Vlasov-Poisson equation and a second 
order equation, respectively. 
In \cite[section 3]{KR2} it is shown that the
force terms of the 1.5PN approximation used here
do agree with the infinte particle limit of the
comparison dynamics used in \cite{KS2}.
  
In contrast to the PN approximation in this paper
in \cite{KS1} and \cite{KS2} only the forces are
expanded  but the main difference is in the
treatment  of the initial data. For the full
particle  model the initial data for the fields
are  supposed to be of ``charged soliton'' type.
One can think of these fields  as generated by
charges  forced to move freely for $-\infty<t\le
0$  with their initial velocity. For the
approximation  this leads to an initial time slip
$t_0$ which the charges need to ``forget'' their
initial data. The initial data of the
approximation is fixed by matching the data of the 
full system at time $t_0$. Therefore the initial
data for the approximation are given only
implicitly, first one has to compute a solution of
the full system over a time span $t_0$. Regarding
the Cauchy problem of Vlasov-Maxwell  system we do
the matching  the other way round. For a given
initial  density one computes the fields of the
approximations and  imposes their values at $t=0$
as initial data on the fields of the full system.
Therefore these initial data are given more
explictly. Even more it is possible to calculate
them  by the values of $f^{\circ, \pm}_0$ and 
$f^{\circ, \pm}_2$  alone, see (\ref{radE-rep}) 
and (\ref{radB-rep}).

Moreover both results in this paper (Theorem
\ref{Hauptsatz} and Theorem \ref{Hauptsatz2}) seem
to be   stronger than the results obtained for the
particle  model.  In \cite{KS2} the error bounds
of the 1.5-PN  approximation in contrast to the
error bounds of the 1-PN-approximation are only
improved  in a certain direction, see
\cite[formulas (3.21) and (3.32)]{KS2}. It seems
reasonable that a matching of  the  initial
conditions at time $t=0$ according to the
treatment of the initial conditions used  here
might improve those bounds.


\subsection{Notation} \label{not-sec}
In the remaining more technical sections of this
paper we shall make use of the following
notations. $B(0, R)$ denotes the closed ball in
$\R^3$ with center at $x=0$ or $p=0$ and radius
$R>0$. We write 
\[ g(x, v, t, c)={\cal O}_{cpt}(c^{-m}), \]
if for all $R>0$ and $T>0$ there is a constant
$M_R>0$ only depending on the basic constants
$r_0, S_0$ and, while dealing with solutions of
(\ref{retRVMC}), $P_1$ and $M_7(\cdot , \cdot,
\cdot)$ such that 
\begin{equation}\label{GOForm}
   |g(x, v, t, c)|\leq M_Rc^{-m}
\end{equation}
for $|x|\le R$, $p\in\R^3$, $t\in [0, T]$, and
$c\geq 1$. Similarly, we write 
\[ g(x, v, t, c)={\cal O}(c^{-m}), \]
if for all $T>0$ there is a constant $M>0$ only
depending on the basic constants such that
(\ref{GOForm}) holds for all $x, p\in\R^3$, $t\in
[0, T]$, and $c\geq 1$. In short, ${\cal O}$ is
global in $x, p$ and $c$ while ${\cal O}_{cpt}$ is
local in $x$ global in $P$ and $c$, and both
symbols are local in $t$. In general, generic
constants only depending on the basic constants
are denoted by $M$. Furthermore, in the following
sections we shall use the notation  
\[ f=f^+-f^-,\quad f^\circ_k=f^{\circ,
  +}_k-f^{\circ,-}_k,\quad f^{\circ}_c=f^{\circ,
  +}_c-f^{\circ, -}_c\text{and}\quad
f_l=f^+_l-f^-_l\] 
 for $k=0, 2$ and $l=0, 2, R$. (Recall that
  $f^\circ_0$ and $f_0$  were used in a different
  way in Subsection \ref{cm-sec}.)   

\section{Proof of Theorem
  \ref{Hauptsatz}}\label{HS-bew} 

\subsection{Estimating
  $E-E^R$}\label{diffE-subsec} 
In Section \ref{repapp-sec} below we will show that
the approximate electric field $E^R$ from
(\ref{rad-def}) admits the following
representation. 
\begin{subequations}
\begin{equation}
  \label{ER-rep}
  E^R=E^R_{ext} +E^R_{int} +E^R_{bound}+{\cal
  O}_{cpt}(c^{-4}) 
\end{equation}
with 
\begin{eqnarray}
  \label{ERext-rep}
  E^R_{ext}(t, x)  & = & \int_{|z|\ge ct}
  \bigg\{-|z|^{-2}\bar{z}
  \big(\rho_0+c^{-2}\rho_2\big)
  -c^{-2}|z|^{-1}\partial_t j_0 
  \\ \nonumber
  && +1/2c^{-2}\bar{z}\partial_t^2\rho_0 +2/3 
  c^{-3}\partial_t\big[E_0(\rho_0^
  ++\rho^-_0)\big]\bigg\}(t,  x+z)\,dz  
  \\ \label{ERint-rep}
  E^R_{int}(t, x) & = & \int_{|z|\le ct}
  \int|z|^{-2} K_1(\bar{z}, c^{-1}p)
  f_R(\hat{t}(z), x+z, p)\,dp\,dz 
  \\ \nonumber
  && +c^{-2}\int_{|z|\le ct} \int|z|^{-1}
  K_2(\bar{z}, c^{-1}p) E_0(\hat{t}(z), x+z)
  (f^+_R+f^-_R)(\hat{t}(z), x+z, p)\,dp\,dz  
  \\ \label{ERbound-rep}
  E^R_{bound}(t, x) & = & (ct)^{-1} \int_{|z|=ct}
  \int K_3(\bar{z}, c^{-1}p)
  (f^\circ_0+c^{-2}f^\circ_2)(x+z, p) \,dp\,ds(z)
  \\ \nonumber 
  && -1/3 tc^{-2}\int_{|z|=ct} \bar{z}
  \bar{z}\cdot\partial_t^2 j_0(0, x+z)\,ds(z) 
  \\ \nonumber 
  && +2/3c^{-3} \int_{|z|=ct} \int(\bar{z}\cdot 
  p)\partial_t j_0(0,x+z, p)\,dp\,ds(z)
\end{eqnarray}
  
\end{subequations}
where the subscripts `ext', `int' and `bound'
refer to the exterior, interior and boundary
integration in $z$. The kernels are given by 
\begin{subequations}
\begin{eqnarray}
  K_1(\bar{z}, \tilde{p}) & = & -\bar{z} +2\bar{z} 
  (\bar{z}\cdot\tilde{p}) -\tilde{p} -3\bar{z}
  (\bar{z}\cdot\tilde{p})^2
  +2(\bar{z}\cdot\tilde{p})\tilde{p}
  +\bar{z}\tilde{p}^2 +4\bar{z}
  (\bar{z}\cdot\tilde{p})^3  
  \nonumber \\
  && -3\tilde{p}(\bar{z}\cdot\tilde{p})^2
  -3\bar{z}(\bar{z}\cdot \tilde{p})\tilde{p}^2
  +3/2\tilde{p}\tilde{p}^2 
  \label{K1-def}\\
  K_2(\bar{z}, \tilde{p}) & = &
  \bar{z}\otimes\bar{z}-1
  -2\bar{z}\otimes\bar{z}(\bar{z}\cdot \tilde{p})
  +\bar{z}\cdot\tilde{p} +\tilde{p}\otimes\bar{z} 
  \label{K2-def} \\
  K_3(\bar{z}, \tilde{p}) & = &
  \bar{z}(\bar{z}\cdot\tilde{p})
  +\tilde{p}(\bar{z}\cdot \tilde{p})
  -\bar{z}(\bar{z}\cdot\tilde{p})^2
  +\bar{z}(\bar{z}\cdot\tilde{p})^3
  -\tilde{p}(\bar{z}\cdot\tilde{p})^2
  -1/2\bar{z}(\bar{z}\cdot\tilde{p})\tilde{p}^2.  
  \nonumber 
\end{eqnarray}
\end{subequations}
We also recall that $\bar{z}=z|z|^{-1}$ and
$\hat{t}(z)=t-c^{-1}|z|$. On the other hand,
according to Section 
\ref{repmax-sec} below, we have
\begin{subequations}
\begin{equation}
  \label{E-rep}
  E=E_{ext} +E_{int}+E_{bound} +{\cal O}(c^{-4}) 
\end{equation}
with 
\begin{eqnarray}
  \label{Eext-rep}
  E_{ext}(t, x) & = & \int_{|z|\ge ct} \bigg\{
  -|z|^{-2}\bar{z}
  \Big[(1+t\partial_t)(\rho_0+c^{-2}\rho_2)
  +(1/2t^2\partial_t^2+1/6t^3\partial_t^3)\rho_0
  \Big] 
  \\[1.5ex] \nonumber
  & & +1/2c^{-2}\bar{z}
  (\partial_t^2+t\partial_t^3)\rho_0
  -c^{-2}|z|^{-1}(\partial_t+t\partial_t)j_0 
  \\[1ex] \nonumber 
  & & +2/3c^{-3}\partial_t[E_0(\rho^+_0
  +\rho^-_0)] \bigg\}(0,
  x+z)\,dz
  \\
  E_{int}(t, x) & = &\int_{|z|\le ct}
  \int|z|^{-2}K_1(\bar{z}, c^{-1}p)f(\hat{t}(z),
  x+z, p)\,dp\,dz 
  \\ \nonumber 
  && +\int_{|z|\le ct}\int|z|^{-1}
  K_2(\bar{z},c^{-1}p)E(\hat{t}(z), x+z)
  (f^++f^-)(\hat{t}(z), x+z, p)\,dp\,dz  
  \label{Eint-rep} \\
  E_{bound}(t, x) & = & E^R_{bound}(t, x).
  \label{Ebound-rep}
\end{eqnarray}

\end{subequations}
In order to verify (\ref{diff-esti}), we fix
constants $R>0$ and $0<T<\min\{\tilde{T},
\hat{T}\}$. For $x\in B(0, R)$  and $t\le T$ we
start by comparing the exterior fields. We obtain
from (\ref{Eext-rep}) and (\ref{ERext-rep}), due
to $|\bar{z}|=1$, and taking into account Theorem
\ref{f0-th}(b)-(c) and Lemma \ref{f2-le}(a)-(b) 
\begin{eqnarray}\label{ext-diff}
   \lefteqn{|E^R_{{ ext}}(t, x)-E_{{ext}}(t, x)|}
   \nonumber 
   \\ & \le & \int_{|z|>ct}
   \bigg\{|z|^{-2}|\rho_0(t,
   x+z)-\big(1+t\partial_t 
   +1/2\,t^2\partial_t^2+1/6\,t^3\partial_t^3\big)\rn\nxz
   |\nonumber 
   \\ & & +c^{-2}|z|^{-2}|\rho_2(t,
   x+z)-(1+t\partial_t)\rho_2(0, x+z)| \nonumber 
   \\ & & +c^{-2}|z|^{-1}\int |p|\,|\partial_t
   f_0(t, x+z, p) 
   -(1+t\partial_t)\partial_t f_0(0, x+z, p)|\,dp
   \nonumber 
   \\ & & +1/2\,c^{-2}|\partial_t^2 \rho_0(t, x+z) 
   -(1+t\partial_t)\partial_t^2 \rn(0, x+z)|
   \nonumber 
   \\ & &
   +2/3\,c^{-3}|\partial_t[\En(\rnp+\rnm)]\txz
   -\partial_t[\En(\rnp+\rnm)]\nxz\bigg\}\,dz 
   \nonumber  
   \\ & \le & M M_2(T) M_1^3(T)\int_{|z|>ct}
   |z|^{-2}t^4  
   {\bf 1}_{B(0,M_1(T))}(x+z)\,dz \nonumber
   \\ & & +c^{-2}M M_4(T) M_3^3(T)\int_{|z|>ct}
   |z|^{-2}t^2 
   {\bf 1}_{B(0, M_3(T))}(x+z)\,dz \nonumber
   \\ & & + c^{-2} M M_2(T) M^4_1(T)
   \int_{|z|>ct}|z|^{-1}t^2 
   {\bf 1}_{B(0, M_1(T))}(x+z)\,dz \nonumber
   \\ & & + c^{-2} M M_2(T)  M^3_1(T)
   \int_{|z|>ct}t^2 
   {\bf 1}_{B(0, M_1(T))}(x+z)\,dz \nonumber 
   \\ & & +c^{-3} M M_2^2(T)
   M_1^3(T)\int_{|z|>ct}t 
   {\bf 1}_{B(0, M_1(T))}(x+z)\,dz \nonumber 
   \\ & \le & M\int_{|z|>ct} \big(t^4|z|^{-2}+t^2
   c^{-2}(|z|^{-2}+\zme+1)+t c^{-3}\big) {\bf
   1}_{B(0, R+M_0)}(z)\,dz 
   \nonumber
   \\ & \le & M_R\,c^{-4};
\end{eqnarray}
where $M_0=\max\{M_1(T), M_3(T), M_5(T)\}$ and we
used that for instance 
$$  t^4\int_{|z|>ct} |z|^{-2}{\bf 1}_{B(0,
   R+M_0)}(x+z)\,dz 
   \le (ct)^{-4}t^4\int_{|z|\le R+M_0}
   |z|^2\,dz\le M_R\,c^{-4}. $$ 

To bound $|E_{{\rm int}}(t, x)-E^R_{{\rm int}}(t,
x)|$, we first claim that 
\begin{equation}\label{ma6}
   |E(t, x)-E_0(t, x)|={\cal O}(c^{-2}). 
\end{equation}
which can be proved analogously to  \cite[Theorem
1.1]{bauku}. (This estimate holds uniformly in
$x\in\R^3$.)  Next we define 
\[ Q^\pm(t)=\sup\,\{|f^\pm(s, x, p)-f^\pm_R(s, x,
p)|: x\in\R^3, p\in\R^3, s\in [0, t]\}. \]  
and $Q(t)=Q^+(t)+Q^-(t)$. Recall that $f^\pm(s, x,
p)=f^\pm_0(s, x, p)=f^\pm_2(s, x, p)=0$  if
$|x|\ge M_0$ or $|p|\ge M_0$. Therefore the
kernels $K_i$, $i+1,2,3$ are bounded on the domain
of the densities by $M(1+M_0^3)$ uniformly in
$c\ge 1$. From (\ref{Eint-rep}),
(\ref{ERint-rep}), (\ref{ma6}), (\ref{IC}), and
$0\le\hat{t}(z)\le t$ for $|z|\le ct$ we obtain 
\begin{eqnarray}\label{int-diff}
   \lefteqn{|E_{{\rm int}}(t, x)-E^R_{{\rm
   int}}(t, x)|}  
   \nonumber \\ 
   & \le & M(1+M_0)^3\int_{|z|\leq
   ct}\int\big[\big( |z|^{-2} 
   +c^{-2}|z|^{-1}|E_0|
   \big)\big(|f^+-f^+_R|+|f^--f^-_R|\big)\big)  
   \nonumber \\
   &&+  c^{-2}\zme|E-E_0|(|f^+|+|f^-|)
   \big]\tdxzp\diffp\dz 
   \nonumber \\ 
   & \le & M (1+M_0^3)M_0^3(1+M_2(T))\,
   Q(t)\int_{|z|\leq ct} (|z|^{-2}+\zme) 
   \,{\bf 1}_{B(0, M_0)}(x+z)\,dz
   \nonumber \\
   & & +\cmv M (1+M_0^3)M_0^3M_6(T)\int_{|z|\leq
   ct} \zme\,{\bf 1}_{B(0, M_0)}(x+z)\,dz 
   \nonumber \\
   & \le & M_R(c^{-4}+Q(t)),
\end{eqnarray}
since for instance
\[ \int_{|z|\leq ct} |z|^{-2}\,{\bf 1}_{B(0,
  M_0)}(x+z)\,dz \le\int_{|z|\leq R+M_0}
|z|^{-2}\,dz\le M_R. \] 
Recalling that the $E_{ bound}(t, x)=E^R_{
  bound}(t, x)$, we can summarize (\ref{E-rep}),
(\ref{ER-rep}), (\ref{ext-diff}), and
(\ref{int-diff}) as 
\begin{subequations}
\begin{equation}\label{abschE}
   |E(t, x)-E^R(t, x)|\le M_R(c^{-4}+Q(t)),
\end{equation}
for $|x|\le R$ and $t\in [0, T]$. 
Using (\ref{BR-rep}) and (\ref{B-rep}) below and
similar arguments as  for the electric fields
yield  
\begin{equation}\label{abschB}
   |B(t, x)-B^R(t, x)|\le M_R(c^{-4}+Q(t)),
\end{equation}
\end{subequations}
for $|x|\le R$ and $t\in [0, T]$. 
\subsection{Estimating
  $f^\pm-f^\pm_R$}\label{difff-subsec}

It remains to estimate
$q^\pm=f^\pm-f^\pm_R$. Using (\ref{RVMC}),
(\ref{rad-def}), (\ref{rrVP}), and (\ref{LVP}), it
is found that 
\begin{eqnarray*}
   \lefteqn{\partial_t q^\pm +\hat{p}\cdot\nabla_x
   \,q^\pm \pm(E+c^{-1}\hat{p}\times
   B)\cdot\nabla_p\, q^\pm}\hspace{2em} 
   \\ & = & -\partial_t
   f^\pm_R-\hat{p}\cdot\nabla_x f^\pm_R
   \mp(E+c^{-1}\hat{p}\times B) \cdot\nabla_p
   f^\pm_R 
   \\ & = & (p -1/2\,c^{-2}p^2\,p -\hat{p})
   \cdot\nabla_x f^\pm_0 
   +c^{-2}(p-\hat{p}) \cdot \nabla_x f^\pm_2 
   \\ & & \pm((E^R-E)+c^{-1}\hat{p} \times(B^R-B))  
   \cdot\nabla_p f^\pm_R 
   \\ & &\pm c^{-1}(p-\hat{p})\times B^R\cdot
   \nabla_p f^\pm_0\pm c^{-4} p\times B_3\cdot
   \nabla_p f^\pm_0 
   \\ & & \mp c^{-2}(E^R-E_0)\cdot\nabla_p f^\pm_2 
   \mp c^{-3}\hat{p}\times B^R\cdot\nabla_p
   f^\pm_2. 
\end{eqnarray*}
If $|p|\le M_0$, then also $|\hat{p}|=(1+c^{-2}
p^2)^{-1/2}|p|\le |p|\le M_0$ uniformly in $c$,
and hence 
\[
\big|\hat{p}-\big(1-1/2\,c^{-2}p^2\big)p\big|\le
Mc^{-4}. \] 
Next we note the straightforward estimates
$|B^R(t, x)|\le M c^{-1}$, $|E^R(t, x)-E_0(t,
x)|\le M c^{-2}$ and $|B_3(t, x)|\le M$ for  $t\in
[0, T]$. In view of the bounds in Theorem
\ref{f0-th}(c) and Lemma \ref{f2-le}(c), thus by
(\ref{abschE}) and (\ref{abschB}), 
\begin{eqnarray}\label{h-est}
   & & |\partial_t q^\pm(t, x, p)
   +\hat{p}\cdot\nabla_x q^\pm(t, x, p) +(E(t, x)
   +c^{-1}\hat{p}\times B(t, x)) \cdot\nabla_p
   \,q^\pm(t, x, p)| 
   \nonumber \\ & & \qquad\qquad\le M_{M_0}(c^{-4}
   +Q(t))=M(c^{-4} +Q(t))
\end{eqnarray}
for $|x|\le M_0$, $|p|\le M_0$, and $t\in [0,
T]$. But in $\{(t, x, p): |x|>M_0\} \cup\{(t, x,
p): |p|>M_0\}$ we have $q^\pm=f^\pm-f^\pm_R=0$ by
the above definition of $M_0>0$. Accordingly,
(\ref{h-est}) is satisfied for all $x\in\R^3$,
$p\in\R^3$, and $t\in [0, T]$.  Now, for any
$x\in\R^3$, $p\in\R^3$ and $t\in [0, T]$ we
compute using (\ref{fpmdarst}) and (\ref{olan}) as
well as (\ref{h-est}) 
\[ \Big|\frac{d}{ds}q^\pm(s, \X, \P)\Big|
=|\partial_t q^\pm +\hat{p}\cdot\nabla_x
q^\pm\pm(E +c^{-1}\hat{p}\times
B)\cdot\nabla_p q^\pm||_{(s, \X, \P)}\le M c^{-4}+
MQ(s), \]
for $s\in[0, T]$. Here the characteristics are
evaluated at $(s; t, x, p)$. Note that 
\[ q^\pm(s, \X, \P)|_{s=0}=(f^\pm -f^\pm_R)(0, \X,
\P)|_{s=0} = c^{-4} (f^{\circ,+}_{c, free}
-f^{\circ,-}_{c, free})(\X, \P)|_{s=0}. \] 
Thus, 
\begin{eqnarray*}
  |q^\pm(t, x, p)| & = & |q^\pm(s, (\X, \P) (s;t,
  x, p))||_{s=t} 
  \\
  & \le &  c^{-4}| f^\pm_{c, free}(0, (\X, \P)
  (0;t, x, p))|+\int_0^t\Big|\frac{d}{ds}q^\pm(s,
  (\X, \P)(s; t, x, p))\Big|\,ds \\
  &\le & M c^{-4}+M\int_0^t Q(s)\,ds.
\end{eqnarray*}
But by the definition of $Q(t)$ it follows that
\[  Q(t)\le M c^{-4}+M\int_0^t Q(s)\,ds \]
for $t\in[0, T]$. Then Gronwall's inequality
implies $Q(t)\le M c^{-4}$ for $t\in[0,
T]$. Inserting this into (\ref{abschE}) 
and (\ref{abschB}) yields the assertion of Theorem
\ref{Hauptsatz} and completes the proof. 
 {\hfill$\Box$}\bigskip
\section{Proof of Theorem
  \ref{Hauptsatz2}}\label{main2-th} 
We recall the following representation of the
electric field $E$ from \cite{cal1}.        
\begin{eqnarray}
  \label{retE-rep}
  E(t, x) & = & \int\!\!\!\! \int|z|^{-2}
  K_T(\bar{z}, \hat{p})\, f(\ast, p) \,dp\,dz  
  \\ \nonumber
  & & + c^{-2}\int\!\!\!\! \int\zme K_S(\bar{z},
  p)(E+c^{-1}\hat{p}\wedge B)(\ast)(f^++f^-)(\ast,
  p)\,dp\,dz 
\end{eqnarray}
where the integrals are to be extended over
$\R^3$, $K_T$ and $K_S$ are given by
(\ref{KT-def}) and (\ref{KS-def}), respectively
and 
\[ (\ast) = (t-c^{-1}|z|, x+z). \]
Now we fix some constants $T<\tilde{T}$ and $R>0.$
Furthermore, we assume that $c\ge 2P_1$ as well as
$|x|\leq R$ and $|t|\leq T$. Define  
\begin{subequations}
\begin{equation}
  \label{pstern-def}
  p^\ast=p^\ast(T)=\max\{P_1, M_1(T), M_3(T)\} 
\end{equation}
and 
\begin{equation}
  \label{rstern-def}
  r^\ast=r^\ast(T, R)= \max\{2(r_0+P_1T)+R,
  M_1(T), M_3(T)\} 
\end{equation}
\end{subequations}
Then $|x+z|\ge r^\ast$ implies
\[ r_0+P_1|t-c^{-1}|z||=r_0+P_1|t-c^{-1}|x+z-x|| 
\le r_0+P_1T+\frac{1}{2}R+\frac{1}{2}|x+z|\leq
|x+z|. \] 
Therefore, $|p|\ge p^\ast $ or $|x+z|\ge r^\ast$
yields $f^\pm(\ast, p)=0$ by Assumption
\ref{retRVMC-ass}(b). This argument shows, that in
(\ref{retE-rep})  we can replace
$\int\!\!\int\,dp\,dz$ by $\int_{|x+z|\le
  r^\ast}\int_{|p|\le  p^\ast}\,dp\,dz$. In other
words, in these integrals we may always assume
that both $|z|$ and $|p|$ are bounded by a bound
depending on $r_0, P_1, R$ and $T$. Since the
$p$-domain is bounded we have 
\begin{subequations}
  \begin{eqnarray}
    K_T(\bar{z}, \hat{p}) & = & K_1(\bar{z},
    c^{-1} p)+{\cal O}( c^{-4}) 
    \label{KT-exp} \\
    K_S(\bar{z}, p) & = & K_2(\bar{z}, c^{-1}
    p)+{\cal O}( c^{-2}) 
    \label{KS-exp}
  \end{eqnarray}
where $K_1$ and $K_2$ are defined in
(\ref{K1-def}) and (\ref{K2-def}),
respectively. Furthermore, 
\begin{equation}
  \label{pd-exp}
  c^{-1} \hat{p} = c^{-1}
  p-c^{-3}\frac{p^2}{2}p+{\cal O}(c^{-5}). 
\end{equation}
Firstly, we have using Assumption
\ref{retRVMC-ass}(b),(c) 
\begin{eqnarray}\label{retB-bound}
  |B(t, x)| & = & c^{-1}\int_{|x+z|\le r^\ast} 
  \zme|\nabla\times j(\ast)|\,dz\\ \nonumber
  & \leq & c^{-1} M_7(T+R+r^\ast, r^\ast,
  p^\ast)\int_{|x+z|\le r^\ast} 
  |z|^{-1}\,dz\int_{|p|\le p^\ast}\,dp = {\cal
  O}_{cpt}(c^{-1}). 
\end{eqnarray}
In the same waywe are able to bound the electric
field
\begin{eqnarray}
  \label{retE-bound}
  |\partial_t^l E(t, x)| & = & \int_{|x+z|\le
   r^\ast} |z|^{-1}\Big|\partial^l_t\nabla\rho
   +c^{-2} \partial_t^{l+1} j\Big|(\ast) \,dz
  \\  \nonumber
  & \le & M_7(T+R+r^\ast, r^\ast,
  p^\ast)\int_{|x+z|\le r^\ast} 
  |z|^{-1}\,dz\int_{|p|\le p^\ast}\,dp = {\cal
  O}_{cpt}(1)
\end{eqnarray}
for $l=0,...,3.$
Hence, using (\ref{retB-bound}) 
\[  c^{-2}\int\!\!\int\zme K_S(\bar{z}, p)
c^{-1} \hat{p}\wedge B(\ast)(f^++f^-)(\ast,
p)\,dp\,dz = {\cal O}_{cpt}( c^{-4}).\] 
For this reason and using (\ref{retE-bound}) we
conclude for the second term in (\ref{retE-rep}) 
\begin{eqnarray*}
  \lefteqn{ c^{-2}\int\!\!\!\!\int|z|^{-1}
    K_S(\bar{z}, p)(E+c^{-1}\hat{p}\wedge
    B)(\ast)(f^++f^-)(\ast, p)\,dp\,dz} \\ 
    & = &  c^{-2}\int\!\!\!\!\int|z|^{-1} K_S(\bar{z},
    p)E\,(\ast)(f^++f^-)(\ast, p)\,dp\,dz +{\cal
    O}_{cpt}( c^{-4})  
    = {\cal O}_{cpt}( c^{-2}).
\end{eqnarray*}
Secondly we shall expand the retardet time. For
every smooth function $\psi$ we have 
\[ \psi(t-c^{-1}|z|) =
\psi(t)-c^{-1}|z|\partial_t\psi(t) +
c^{-2}\frac{|z|^2}{2}\partial_t^2\psi(t) 
-c^{-3}\frac{|z|^3}{6}\partial_t^3\psi(t) +
c^{-4}\frac{|z|^4}{24}\partial_t^4\psi(\xi)\] 
with some $t-c^{-1}|z|<\xi<t$. Using Assumption
\ref{retRVMC-ass}(b),(c) and (\ref{retE-bound}) we
have  
\begin{equation}\label{retET-exp}
  \int_{|x+z|\le r^\ast}|z|^{-2+l} \int_{|p|\le
  p^\ast} K_T(\bar{z}, \hat{p})
  \partial_t^l f(\xi(t, x, z, p), x+z, p) \,dp\,dz
  = {\cal O}_{cpt}(1) 
\end{equation}
for $l=0, \ldots, 4$ as well as
\begin{equation}\label{retES-exp}
  \int_{|x+z|\le r^\ast}|z|^{-1+l}\int_{|p|\le
  p^\ast} 
  K_L(\bar{z},
  p)\partial_t^l\big[E(f^++f^-)\big](\xi(t, x, z,
  p), x+z, p)\,dp\,dz = {\cal O}_{cpt}(1)
\end{equation}
\end{subequations}
for $l=0, 1, 2$ and any choice of  $t-c^{-1}|z|\le
\xi(t, x, z, p)\le t$ 
Now from (\ref{KT-exp})-(\ref{retES-exp}) we can
expand the electric field in  powers of
$c^{-1}$. To order zero we have 
\begin{eqnarray}
  \label{retE0}
  E(t, x) & = &  -\int_{|x+z|\le
  r^\ast}|z|^{-2}\bar{z} \rho(t, x+z)\,dz + 
  {\cal O}_{cpt}(c^{-1})
  \nonumber \\
  & = &  -\int|z|^{-2}\bar{z} \rho(t, x+z)\,dz + 
  {\cal O}_{cpt}(c^{-1}).
\end{eqnarray}
Here we employed that $f^\pm(t, x+z, p)=0$ if
$|x+z|\ge r^\dagger$ with  
\begin{equation}
  \label{rdagger-def}
  r^\dagger= r_0+P_1T,
\end{equation}
 see Assumption \ref{retRVMC-ass}(b) and use
 $r^\ast>r^\dagger$, see (\ref{rstern-def}). In
 the first order we obtain  two terms, the first
 coming from the expansion of the kernel $K_T$ and
 the second from the  expansion  of the retardet
 time, 
\[
 c^{-1}\int\!\!\!\!\int|z|^{-2}(2\bar{z}(\bar{z}\cdot
 p)-p)f(t, x+z, p)\,dp\,dz \] 
and
\[ c^{-1}\int|z|^{-1}\bar{z}\partial_t\rho(t, x+z)\,dz. \]
Here and in the sequel we shall treat terms
containig  time derivatives in the following
way. Firstly, we replace $\partial_t f^\pm$ by
$-\hat{p}\cdot\nabla_x f^\pm
\mp(E+c^{-1}\hat{p}\wedge B)\cdot \nabla_p f^\pm$
using the Vlasov equation. Secondly we do an
integration by parts utilizing  
\[ (E+c^{-1}\hat{p}\wedge B)\cdot \nabla_p f^\pm
=\nabla_p\cdot 
\big[(E+c^{-1}\hat{p}\wedge B)f^\pm\big]. \]
For the second term we therefore obtain
\[ -c^{-1}
\int\!\!\!\!\int|z|^{-2}
(2\bar{z}(\bar{z}\cdot\hat{p})-\hat{p})f(t, x+z,
p)\,dp\,dz.\]  
Hence, using (\ref{pd-exp}),  the contribution in
first order vanishes and  in (\ref{retE0}) we can
replace ${\cal O}_{cpt}(c^{-1})$ by ${\cal
  O}_{cpt}( c^{-2})$,  but we have an additional
term  occuring in the third order, namely 
\begin{equation}
  \label{3o-add}
  c^{-3}\int\!\!\!\!\int|z|^{-2}
  p^2(\bar{z}(\bar{z}\cdot p)-\frac{1}{2}p)\,f(t,
  x+z, p) \,dp\,dz.
\end{equation}
In the next orders we will simply replace
$\hat{p}$ by $p$ according to (\ref{pd-exp})
without comment. For the second order we start
with the terms with two time derivatives;  
\begin{subequations}
  \begin{eqnarray}
    \label{secotwoder}
    \lefteqn{-
    c^{-2}\frac{1}{2}\int\bar{z}\partial_t^2
    \rho(t, x+z)dz} 
    \\ \nonumber 
    & = & - c^{-2}\frac{1}{2}\int\!\!\!\!\int
    |z|^{-1}(-\bar{z}(\bar{z}\cdot p) +p)
    \partial_t  f(t, x+z, p) \,dp\,dz +{\cal
    O}(c^{-4}).  
  \end{eqnarray}
Now we collect and rewrite the terms with one time
derivative, including the term coming from
(\ref{secotwoder}), 
\begin{eqnarray}
  \label{secooneder}
  \lefteqn{- c^{-2}\int\!\!\!\! \int|z|^{-1}
 \bigg(2\bar{z} (\bar{z}\cdot p)-p+\frac{1}{2}
 (-\bar{z}(\bar{z}\cdot p) +p)\bigg)\partial_t
 f(t, x+z, p) \,dp\,dz}
 \nonumber \\
  & = & -\int\!\!\!\! \int|z|^{-2}
 \bigg(-\frac{9}{2}\bar{z} (\bar{z}\cdot p)^2
  +2 (\bar{z}\cdot p)\,p
 +\frac{3}{2}\bar{z}\,p^2\bigg) f(t, x+z, p)
 \,dp\,dz 
  \label{secozeroder}\\
  & &
  -\int\!\!\!\!  \int|z|^{-1}
  \bigg(\frac{3}{2}\bar{z}\otimes\bar{z}
  -\frac{1}{2}\bigg) 
  E(t, x+z)(f^+ +f^-)(t, x+z, p) \,dp\,dz + {\cal
  O}( c^{-4}). 
  \nonumber
\end{eqnarray}
Together with  the remaining terms coming from the
expansion of $K_T$ we therefore obtain 
\begin{eqnarray}
  \label{retf2}
  E(t, x) & = &
  \int\!\!\!\!\int|z|^{-2}\bigg\{-\bar{z}+
  c^{-2}\Big(\frac{3}{2}\bar{z}(\bar{z}\cdot 
  p)^2-\frac{1}{2}\bar{z} p^2\Big)\bigg\}f(t,
  x+z, p)\,dp\,dz
  \\  \nonumber 
  & & -
  c^{-2}\frac{1}{2}\int\!\!\!\!\int|z|^{-1}
  \bigg\{\bar{z}\otimes\bar{z}+1\bigg\}E(t, x+z)  
  (f^+ +f^-)(t, x+z, p)\,dp\,dz + {\cal
  O}_{cpt}(c^{-3}). 
\end{eqnarray}
\end{subequations}
At last we turn to the third order and, following
the usual route, first treat the term  with three
time derivatives, 
\begin{subequations}
  \begin{eqnarray}
    \label{thirdotthreeder}
    c^{-3}\frac{1}{6} \int z\partial_t^3 \rho(t,
    x+z)\,dz & = &  
    c^{-3}\frac{1}{6} \int\!\!\!\!\int
    p\partial_t^2 f(t, x+z, p)\,dp\,dz +{\cal
    O}(c^{-5}).  
  \end{eqnarray}
Regarding terms containing second time derivatives
including the term coming from
(\ref{thirdotthreeder}) we have 
\begin{eqnarray}
  \label{thirdotwoder}
  c^{-3} \lefteqn{\int\!\!\!\!
  \int\bigg\{\bar{z}(\bar{z}\cdot 
  p)-\frac{1}{2}p +\frac{1}{6}p\bigg\}
  \partial_t^2 f(t, x+z, p)\,dp\,dz} 
  \nonumber \\ \nonumber &&
  =c^{-3}\int\!\!\!\!\int|z|^{-1}
  \bigg\{-2\bar{z}(\bar{z}\cdot p)^2
  +(\bar{z}\cdot p)p
  +\bar{z}\,p^2\bigg\}\partial_t f(t, x+z, p)
  \,dp\,dz 
  \\ & & 
  +c^{-3}\int\!\!\!\! \int\Bigg
  \{\bar{z}\otimes\bar{z} -\frac{1}{3}\bigg\}
  \partial_t\Big[E(f^+ +f^-)\Big](t,
  x+z, p)\,dp\,dz+{\cal O}( c^{-5})
\end{eqnarray}
Now we collect those term which contain exactly
one time derivative including those coming from
(\ref{thirdotwoder}), 
\begin{gather}
  c^{-3}\int\!\!\!\!\int|z|^{-1}\bigg\{-(-3\bar{z}(\bar{z}\cdot
  p)^2+2(\bar{z}\cdot 
  p)\,p+\bar{z}\,p^2)-2\bar{z}(\bar{z}\cdot
  p)^2+(\bar{z}\cdot p)\,p+\bar{z}\,p^2 
  \bigg\} 
  \nonumber \\
   \hspace{-12em}\partial_t f(t, x+z, p)\,dp\,dz
  \nonumber \\
  +c^{-3}\int\!\!\!\!\int\bigg\{-(\bar{z}\otimes\bar{z}-1)
  +\bar{z}\otimes\bar{z} 
  -\frac{1}{3}\bigg\}
  \partial_t\big[E(f^++f^-)\big](t, x+z,
  p)\,dp\,dz 
  \nonumber \\
  =c^{-3}\int\!\!\!\!\int|z|^{-2}
  \bigg\{-4\bar{z}(\bar{z}\cdot p)^3
  +3(\bar{z}\cdot p)^2\,p 
  +2(\bar{z}\cdot p)\,p^2\bar{z}
  -p^2\,p 
  \bigg\}f(t, x+z, p)\,dp\,dz
  \nonumber \\
  +c^{-3}\int\!\!\!\!\int\bigg\{2(\bar{z}\cdot
  p)\bar{z}\otimes\bar{z} -p\otimes\bar{z} 
  -\bar{z}\cdot p\bigg\}E(t, x+z)(f^++f^-)(t, x+z,
  p)\,dp\,dz 
  \nonumber \\
  +c^{-3}\frac{2}{3}\partial_t\int\!\!\!\!\int
  E(t, x+z)(f^++f^-)(t, x+z, p)\,dp\,dz+{\cal
    O}(c^{-5}).
  \label{thirdooneder} 
\end{gather}
Note that we do not touch upon the time derivative
in the term of the last line. This term is
responsible for radiation effects. Collecting all
terms without time derivative coming from the
third order in the expansion of $K_T$, the first
order of the expansion of $K_S$,
(\ref{thirdooneder}) and (\ref{3o-add}) we note
that all these terms cancel exactly. Thus, the
only remaining contribution in third order is the
radiation term   
\[ \partial_t\int\!\!\!\!\int E(t,
x+z)(f^++f^-)(t, x+z, p)\,dp\,dz  
  = \partial_t\int\!\!\!\!\int E(t, z)(f^++f^-)(t,
  z, p)\,dp\,dz=:\tilde{D}(t) \] 
Next we try to recast this term making it
  comparable to $\Dd$.  Note that using
the bounds on  $f^\pm$, see Assumption
  \ref{retRVMC-ass}(c),  and the bounds on the
  support of $f^\pm$, with some
  $t-c^{-1}|z|<\xi(t, x, z)<t$  
  \begin{eqnarray*}
    \partial_t E(t, x) & = &
    -\partial_t\int(\nabla \rho+ c^{-2}\partial_t 
    j)(\ast)\frac{dz}{|z|} \\
    & = & -\int_{|x+z|\le r^\ast}(\partial_t\nabla
    \rho+ c^{-2}\partial_t^2 
    j)(\ast)\frac{dz}{|z|} \\
    & = & -\int_{|x+z|\le r^\ast}\partial_t\nabla 
    \rho(\ast)\frac{dz}{|z|}+{\cal O}_{cpt}(
    c^{-2}) \\ 
    & = & -\int_{|x+z|\le r^\ast}\partial_t\nabla 
    \rho(t, x+z)\frac{dz}{|z|}\\ 
    & &  +c^{-1}\int_{|x+z|\le
    r^\ast}\partial_t^2\nabla 
    \rho(\xi(t, x, z),x+z)\,dz + {\cal O}_{cpt}(
    c^{-2})\\ 
    & = & -\int|z|^{-2}\bar{z}\partial_t\rho(t,
    x+z)\,dz+{\cal O}_{cpt}(c^{-1}) 
  \end{eqnarray*}
Furthermore, we recall (\ref{retE0}) and emphasize
that  the continuity equation (\ref{cont-eq})
holds for both species seperately. Using these
ingredients we compute 
\begin{eqnarray*}
  \tilde{D}(t) & = & \partial_t\int_{|z|\le
  r^\dagger}\!\!\int E(t, 
  z)(f^++f^-)(t, z, p)\,dp\,dz \\
  & = & \int_{|z|\le
  r^\dagger}\int
  |y-z|^{-2}(\overline{y-z})\bigg[-\partial_t(\rho^+-\rho^-)(t,  
  y)(\rho^++\rho^-)(t, z)\\
  & & \hspace{2em}-(\rho^+-\rho^-)(t,
  y)\partial_t(\rho^++\rho^-)(t, z) 
  \bigg]\,dy\,dz
  +{\cal O}(c^{-1}) \\
  & = &
  -2\int\!\!\!\!\int|y-z|^{-2}(\overline{y-z})\bigg[\partial_t\rho^+(t, 
  y)\rho^-(t, z)-\partial_t\rho^-(t, y)\rho^+(t,
  z)\bigg]\,dy\,dz  +{\cal O}(c^{-1})\\ 
  & = & 2\int\big(\tilde{H}{}^+(t, z) j^-(t,
  z)-\tilde{H}{}^-(t, 
  z) j^+(t, z)\big)\,dz+{\cal O}(c^{-1}),
\end{eqnarray*}
where
\begin{equation}
  \nonumber
  \tilde{H}{}^\pm(t, x) =
  \oint\zmd(-3\bar{z}\otimes\bar{z}+1)\rho^\pm(t,
  x+z)\,dz  
\end{equation}
and $\rho^\pm$ and $ j^\pm$ are defined in the
obvious way. 
Summarizing, we have 
\begin{eqnarray}
  \label{retE3}
   E(t, x) & = &
  \int\!\!\!\!\int|z|^{-2}\bigg\{-\bar{z} +
  c^{-2}\Big(\frac{3}{2}\bar{z}(\bar{z}\cdot 
  p)^2-\frac{1}{2}\bar{z} p^2\Big)\bigg\}f(t,
  x+z, p)\,dp\,dz
  \\  \nonumber 
  & & -
  c^{-2}\frac{1}{2}\int\!\!\!\!\int|z|^{-1}\bigg\{\bar{z}\otimes\bar{z}+1\bigg\}E
  (t, x+z) 
  (f^++f^-)(t, x+z, p)\,dp\,dz 
  \\ \nonumber 
  & & +c^{-3}\frac{4}{3}\int\big(\tilde{H}{}^+(t, z) j^-(t, z)-\tilde{H}{}^-(t,
  z) j^+(t, z)\big)\,dz+ {\cal O}( c^{-4}). 
\end{eqnarray}
\end{subequations}

Analogous computations lead to 
\begin{subequations}
\begin{eqnarray}
  \label{retB0}
  \lefteqn{B(t, x)  =  {\cal O}_{cpt}(c^{-1})}
  \\ \label{retB2}
  & = &
  c^{-1}\int\!\!\!\!\int|z|^{-2}\bar{z}\wedge p \,
  f(t, x+z, p)\,dp,dz+ 
  {\cal O}_{cpt}(c^{-3})
  \\ \nonumber\label{retB3}
  & = &  c^{-1}\int\!\!\!\!\int|z|^{-2}\bigg\{\bar{z}\wedge p
  - c^{-2}\frac{3}{2}\bar{z}\wedge p(\bar{z}\cdot
  p)^2 \bigg\}  f(t, x+z, 
  p)\,dp,dz
  \\ \nonumber 
  & &
  +\frac{c^{-3}}{2}\int\!\!\!\!\int|z|^{-1}\bigg\{(\bar{z}\cdot 
  p)\bar{z}\wedge(\cdots)+(\bar{z}\wedge
  p)\otimes\bar{z}\bigg\}E(t, 
  x+z)(f^++f^-)(t, x+z, p)\,dp\,dz
  \\ \nonumber 
  & & +c^{-3}\int\bar{z}\wedge\big(\tilde{H}{}^+(t,
  x+z)j^-(t, x+z)-\tilde{H}{}^-(t, x+z) j^+(t,
  x+z)\big),dz 
  \\
  & &  +{\cal O}_{cpt}( c^{-4}).
\end{eqnarray}
\end{subequations}
Now we have to prove the error estimates step by
step, in fact  for proving the 1.5PN approximation
we need to know that $E=E_0+{\cal O}_{cpt}(
c^{-2})$. 
\begin{lemma}[Newton approximation]\label{new-le}
For all $0<T<\tilde{T}$ and $R>0$ there are
constants $M(T)$ and $M(T, R)$ such that  
\begin{eqnarray*}
  |f^\pm(t, x, p)-f^\pm_0(t, x, p)| & \le &  M(T)
   c^{-2}\quad\hspace{0.55em} (x\in\R^3), 
   \nonumber \\
   |E(t, x)-E_0(t, x)| & \le & M(T, R)\,c^{-2}
   \quad  (|x|\le R), \nonumber  \\
   |B(t, x)-c^{-1} B_1(t, x)| & \le & M(T, R)\,c^{-2}
   \quad  (|x|\le R), \nonumber
\end{eqnarray*}
for all $p\in\R^3$, $t\in [0, T]$, and $c\geq
P_1$.
\end{lemma}
The proof will not be carried out here because
with some rather obvious modifications it closely
follows the lines of Section \ref{HS-bew}  using
the representations (\ref{retE0}), (\ref{retB0})
and  
\begin{equation}
  \label{EradE0-est}
  E^R(t, x)  = E_0(t, x)+{\cal
  O}(c^{-2})\quad\text{and}\quad 
  B^R(t, x)  = c^{-1} B_1(t, x)+  {\cal
  O}(c^{-2}), 
\end{equation}
the last two estimates  are an easy consequence of
(\ref{rad-def}) and the bounds of Proposition
\ref{f0-th} and Lemma \ref{f2-le}.  
\begin{remark}
  If we replace $(f^\pm_0, E_0)$  with a solution
  of (\ref{VP}) the Newton approximation can be
  obtained in the same way. 
\end{remark}
 
As a last step in the proof of Theorem
\ref{Hauptsatz2} we have to provide an estimate of the 
differences of the third order terms. Using the
Calderon-Zygmund inequality \cite[Thm. 4.31]{ada}
and (\ref{H-def}) we have the estimate 
\begin{subequations}
  \begin{eqnarray}
    \nonumber 
    \bigg|\bigg|\oint |z|^{-3}H(z)
    \rho^\pm_{(0)}(t, \cdot +z)
    \,dz\bigg|\bigg|_{L^2}& \leq & C_{CZ}
    ||\rho^\pm_{(0)}(t, \cdot)||_{L^2} 
    \\ \nonumber
  \end{eqnarray}
with a certain constant $C_{CZ}$.
Hence
\begin{eqnarray}\label{diffD-est}
  \nonumber
  \lefteqn{\bigg|\int \big(\tilde{H}^+
  j^--\tilde{H}^-j^+-H^+j^-_0+H^-j^+_0\big)(t,
  z)\,dz\bigg|} 
  \\ \nonumber 
  & \leq & ||(\tilde{H}{}^+-H^+)(t,
  \cdot)||_{L^2}\cdot ||j^-(t, 
  \cdot)||_{L^2} 
  + || H^+(t, \cdot)||_{L^2}\cdot
  ||(j^--j^-_0)(t,\cdot)||_{L^2}  
  \\ \nonumber 
  & &  + ||(\tilde{H}{}^--H^-)(t,
  \cdot)||_{L^2}\cdot ||j^+(t, 
  \cdot)||_{L^2}
  + || H^-(t, \cdot)||_{L^2}\cdot
  ||(j^+-j^+_0)(t,\cdot)||_{L^2}  
  \\ \nonumber 
  & \leq & C_{CZ}\bigg(||(\rho^+-\rho_0^+)(t,
  \cdot)||_{L^2}\cdot ||j^-(t, 
  \cdot)||_{L^2}
  + || \rho^+(t, \cdot)||_{L^2}\cdot
  ||(j^--j^-_0)(t,\cdot)||_{L^2}  
  \\ \nonumber 
  & &  + ||(\rho^--\rho_0^-)(t,
  \cdot)||_{L^2}\cdot ||j^+(t, 
  \cdot)||_{L^2}
  + || \rho_0^-(t, \cdot)||_{L^2}\cdot 
  ||(j^+-j^+_0)(t,\cdot)||_{L^2}\bigg)
  \\ \nonumber 
  & \leq & C_{CZ} \,Q^0(t)\, M
  \,(r^\dagger)^3\,(p^\ast)^7  
  \\ 
  & \leq & M Q^0(t)
\end{eqnarray}
where
\[ Q^0(t) := \max\big\{ |f^\pm(t, x, p)-f^\pm_0(t,
x, p)|\;:\;  x\in\R^3, p\in \R^3, \pm\in\{+,
-\}\big\} \] 

Here we used (\ref{rdagger-def}) and Assumption
\ref{retRVMC-ass}(a) in combination with
(\ref{retfini-def}) and (\ref{finf-est}). 
\end{subequations}
In view of the formulas (\ref{radE-rep}),
(\ref{retE3}), Lemma \ref{new-le},
(\ref{EradE0-est}), (\ref{diffD-est}) 
and proceeding analogously to
Section \ref{diffE-subsec}   we conclude the
estimate 
\begin{equation}
  \nonumber
  |E(t, x)-E^R(t, x)|\le M_R( c^{-4}+Q(t)))
\end{equation}
for all $|t|\le T, |x|\le R$ with a constant $M_R$ independent of $c$
where 
\[ Q(t) = \max\big\{ |f^\pm(t, x, p)-f^\pm_R(t, x, p)|\;:\;
x\in\R^3, p\in \R^3, \pm\in\{+, -\}\big\} \]
Employing the formulas (\ref{radB-rep}) and (\ref{retB3}) together with
an estimate corresponding to (\ref{diffD-est}) we conclude
\begin{equation}
  \nonumber
  |B(t, x)-B^R(t, x)|\le M_R( c^{-4}+Q(t)))
\end{equation}
for all $|t|\le T, |x|\le R$ with a constant $M_R$ independent of $c$.
Proceeding analogously to Section \ref{difff-subsec} finishes the
proof of Theorem \ref{Hauptsatz2}.{\hfill$\Box$}\bigskip
\section{Appendix}\label{append} 
\subsection{Representation of the approximation fields $E^R$ and $B^R$}
\label{repapp-sec}

Here we will derive the representation formula
(\ref{ER-rep}) for the approximate field $E^R$
from (\ref{rad-def}). Since the calculations for
the electric and the magnetic field are quite
similar, we will only analyse in detail the
electric field and simply state the result for its
magnetic counterpart. Actually, the representation
of the electric field exhibits more difficulties
than the representation of the magnetic field, due
to the presence of the radiation term. 

From (\ref{rad-def}) we recall $E^R=E_0+c^{-2}E_2+
(2/3)c^{-3} D^{[3]}$, where 
\begin{subequations}
\begin{eqnarray}
   E_0(t, x) & = & -\int |z|^{-2}\bar{z}\,\rho_0(t, x+z)\,dz, 
   \label{ma1} \\
   E_2(t, x) & = & \frac{1}{2}\int\bar{z}\,\partial_t^2\rho_0(t, x+z)\,dz
   -\int |z|^{-1}\partial_t j_0(t, x+z)\,dz
   \nonumber \\
   & & -\int |z|^{-2}\bar{z}\,\rho_2(t, x+z)\,dz,\qquad
   \label{ma2} \\
   D^{[3]}(t) & = & \dot{D}^{[2]}
   =\partial_t\int\!\!\!\! \int E_0(t, x)(f^+_0+f^-_0)(t, x, p)\,dp\,dx 
   \label{ma3}
\end{eqnarray}
\end{subequations}
cf.~(\ref{rrVP}) and (\ref{LVP}). 
Firstly, we write
\begin{equation} \nonumber
  D^{[3]}(t) =\partial_t\int\!\!\!\! \int E_0(t, x+z)(f^+_0+f^-_0)(t, x+z, p)\,dp\,dz,\qquad x\in \R^3.
\end{equation}
Secondly, we split the domain of integration in
$\{|z|>ct\}$ and $\{|z|\leq ct\}$; note that the
exterior part exactly gives $E^R_{ext}$ in
(\ref{ER-rep}). To handle the interior part
$\{|z|\leq ct\}$, in the sequel denoted by
$\widetilde{E^R_{int}}$, we fix $R>0$ and
$0<T<\min\{\tilde{T}, \hat{T}\}$  and put  
\[ r^\ddagger=R+M_0\qquad \text{and}\qquad p^\dagger=M_0, \] 
recall that $M_0=\max\{M_1(T), M_3(T), M_5(T)\}$;
$r^\ddagger$ and $p^\dagger$ are only depending on
our basis constants $r_0$, $S_0$, see
(\ref{dat-def}), and of course of $R$. For $|x|\le
R$ and $0\le t\le T$, if $|z|\ge r^\ddagger$ or
$|p|\ge p^\dagger$, than $f^\pm_0(\tau, x+z,
p)=f^\pm_2(\tau, x+z, p)=0$ for all $0\le\tau\le
T$ by Theorem \ref{f0-th}(b) and Lemma
\ref{f2-le}(a). This argument shows that we can
replace $\int_{|z|\le ct}\int\,dp\,dz$ by
$\int_{|z|\le \min\{ct, r^\ddagger\}}\int_{|p|\le
  p^\dagger}\,dp\,dz$ in the integrals defining
the interior part. In other words, we may always
assume that both $|z|$ and $|p|$ are bounded with
a bound only depending on the basic constants and
$R$, but not on  $c$. Next we expand the densities
w.r.t.~$t$ about the retarded time
$\hat{t}(z):=t-c^{-1}|z| $ and obtain e.g.   
\begin{subequations}
\begin{eqnarray}
  \label{ma1-exp}
   \lefteqn{-\int_{|z|\le ct} |z|^{-2}\bar{z}\,\rho_0(t, x+z)\,dz} 
  \\ \nonumber
  &=& -\int_{|z|\le ct}
  |z|^{-2}\bar{z}\,\Big((1+c^{-1}|z|\partial_t+1/2c^{-2}|z|^2\partial_t^2+
  1/6c^{-3}|z|^3\partial_t^3)
  \\ \nonumber 
  & & \hspace{15em}\rho_0(\hat{t}(z), x+z)+|z|^4{\cal O}(c^{-4})\Big)\,dz
  \\ \nonumber
  &=& -\int_{|z|\le ct}
  |z|^{-2}\bar{z}\,(1+c^{-1}|z|\partial_t+1/2c^{-2}|z|^2\partial_t^2+
  1/6c^{-3}\partial_t^3)\,\rho_0(\hat{t}(z), x+z)\,dz
  \\ \nonumber 
  & & +{\cal O}_{cpt}(c^{-4})
\end{eqnarray}
where Theorem \ref{f0-th}(c) was utilized and hence  $M_2(T)$ enters the bounds on ${\cal O}(c^{-4})$ and ${\cal O}_{cpt}(c^{-4})$.
In the same manner we expand the terms in (\ref{ma2}) up to first order and the term from (\ref{ma3}) up to zeroth order,
employing Theorem \ref{f0-th}(c) and Lemma \ref{f2-le}(b), therefore also $M_4(T)$ enters the bounds.
\begin{eqnarray}
  \label{ma2-exp}
  \lefteqn{c^{-2}\int_{|z|\le ct}\big\{1/2\bar{z}\partial_t^2\rho_0-|z|^{-1}\partial_t j_0-|z|^{-2}\bar{z}\rho_2\big\}(t, x+z)\,dz}
   \\ \nonumber
   & = & c^{-2}\int_{|z|\le ct}\big\{\Big(1+c^{-1}|z|\partial_t\Big)\Big(1/2\bar{z}\partial_t^2\rho_0-|z|^{-1}\partial_t
   j_0-|z|^{-2}\bar{z}\rho_2\Big)\big\}(\hat{t}(z), x+z)\,dz
   \\ \nonumber 
   & & +{\cal O}_{cpt}(c^{-4})
   \\ \label{ma3-exp}
   \lefteqn{2/3 c^{-3}\int_{|z|\le ct} \partial_t\big[E_0(\rho^+_0+\rho^-_0)\big](t, x+z)\,dz}
   \\ \nonumber
   & = & 2/3 c^{-3}\int_{|z|\le ct} \partial_t\big[E_0(\rho^++\rho^-)\big](\hat{t}(z), x+z)\,dz+{\cal O}_{cpt}(c^{-4}).
\end{eqnarray}

\end{subequations}
Next we sort the terms according to their orders in $c^{-1}$. To zeroth order we have
\begin{equation}
  \label{0-or}
  \widetilde{E^R_{int}}(t, x) = -\int_{|z|\le ct}|z|^{-2}\bar{z}\rho_0(\hat{t}(z), x+z)\,dz+{\cal O}(c^{-1}).
\end{equation}
In the first order we have the contribution  $-c^{-1}\int_{|z|\le ct}|z|^{-1}\bar{z}\partial_t\rho_0(\hat{t}(z), x+p)\,dz$, which gives,
using that $\partial_t\rho_0+\nabla\cdot j_0=0$ also holds true for (\ref{rrVP}) and further integration by parts,
\begin{subequations}
\begin{eqnarray}
 \lefteqn{c^{-1}\int_{|z|\le ct}|z|^{-1}\bar{z}\nabla_x\cdot j_0(\ldots)\,dz}
 \nonumber \\
 & = & c^{-1}\int_{|z|\le ct}|z|^{-1}(\nabla_z+c^{-1}\bar{z}\partial_t)\cdot\big[j_0(\hat{t}(z), x+z)\big]\,dz
 \nonumber \\
 & = & c^{-1}(ct)^{-1}\int_{|z|=ct}\bar{z}\bar{z}\cdot j_0(0, x+z)\,ds(z)
      -c^{-1}\int_{|z|\le ct}|z|^{-2}(-2\bar{z}\bar{z}\cdot j_0+j_0)(\ldots)\,dz
 \nonumber \\
 && +c^{-2}\int_{|z|\le ct}|z|^{-1}\bar{z}\bar{z}\cdot\partial_t j_0(\ldots)\,dz,
\label{r1-or}
\end{eqnarray}
where $(\ldots)=(\hat{t}(z), x+z)$.
Hence, the first order term of $\widetilde{E^R_{int}}$ can be written as
\begin{equation}
  \label{1-or}
  c^{-1}(ct)^{-1}\int_{|z|=ct}\bar{z}\bar{z}\cdot j_0(0, x+z)\,ds(z)
      -c^{-1}\int_{|z|\le ct}|z|^{-2}(-2\bar{z}\bar{z}\cdot j_0+j_0)(\ldots)\dz
\end{equation}
\end{subequations}
To continue we collect the terms of second order from (\ref{ma2-exp}), (\ref{ma1-exp}), (\ref{r1-or}).
\begin{gather}
  \nonumber
  c ^{-2}\int_{|z|\le ct} \Big(1/2\bar{z}\partial_t^2 
  \rho_0-|z|^{-1}\partial_t j_0-|z|^{-2}\bar{z}\rho_2-1/2\bar{z}\partial_t^2\rho_0+|z|^{-1}\bar{z}\bar{z}\cdot\partial_t j_0\Big)(\ldots)\,dz
\end{gather}
Since the terms containing second derivatives cancel each other we start with the first order time
derivatives.  Utilizing (\ref{rrVP}) and integration by parts we calculate
\begin{subequations}
\begin{eqnarray}
  \label{r2-or}
  \nonumber
  \lefteqn{c^{-2}\int_{|z|\le ct}|z|^{-1}(\bar{z}\bar{z}\cdot\partial_t j_0-\partial_t j_0)(\ldots)\,dz}
  \\ \nonumber
   & = & c^{-2}\int_{|z|\le ct}|z|^{-1}(\bar{z}\otimes\bar{z}-1)(E_0+c^{-3}2/3\Dd)(\rho_0^++\rho_0^-)(\ldots)\,dz
   \\ \nonumber
   & & +c^{-2}\int_{|z|\le
  ct}\int|z|^{-2}f_0(\ldots, p)\big(-3\bar{z}(\bar{z}\cdot p)^2+2\bar{z}\cdot p
  p+\bar{z}p^2\big)\,dp\,dz
  \\ \nonumber
  & & -c^{-2}(ct)^{-1}\int_{|z|=ct}\int(\bar{z}\cdot p)(\bar{z}\bar{z}\cdot p-p) f^\circ_0(x+z, p)\,dp\,dz
  \\
  & & -c^{-3}\int_{|z|\le
  ct}\int|z|^{-1}(\bar{z}\cdot
  p)(\bar{z}\bar{z}\cdot p-p)\partial_t
  f_0(\ldots, p)\,dp\,dz.
\end{eqnarray}
Note that since the bounds in Theorem \ref{f0-th}(b)(c) also imply that
$2/3c^{-5}\int_{|z|\le ct}|z|^{-1}(\bar{z}\otimes\bar{z}-1)\Dd(\rho_0^++\rho_0^-)(\ldots)\,dz={\cal O}_{cpt}(c^{-5})$ this term
can be dropped. Using the same argument all terms containing $\Dd$ appearing in the sequel are at least ${\cal
  O}_{cpt}(c^{-5})$ and hence will be  dropped without comment.
Thus, the contribution of second order  in $\widetilde{E^R_{int}}$ is
\begin{gather}
  c^{-2}\int_{|z|\le
  ct}|z|^{-1}(\bar{z}\otimes\bar{z}-1)E_0(\ldots)(f_0^+
  +f_0^-)(\ldots, p) \,dp\,dz 
  \nonumber \\
  + c^{-2}\int_{|z|\le ct}
  |z|^{-2}\bigg\{f_0(\ldots, p)\big(-3\bar{z}(\bar{z}\cdot p)^2+2\bar{z}\cdot p
  p+\bar{z}p^2\big)-\bar{z}f_2(\dots, p)\Big\} \,dp\,dz
  \nonumber \\ 
  -c^{-2}(ct)^{-1}\int_{|z|=ct}\int(\bar{z}\cdot p)(\bar{z}\bar{z}\cdot p-p)\fon(x+z, p)\,dp\,dz.
  \label{2-or}
\end{gather}
\end{subequations}
We continue by collecting the terms of third order from (\ref{ma2-exp}), (\ref{ma3-exp}), (\ref{ma1-exp}) and
(\ref{r2-or}).
\begin{subequations}
\begin{gather}
  \nonumber
  c^{-3}\int_{|z|\le ct}
  \Big\{(-1/6+1/2)z\partial_t^3\rho_0-\partial_t^2j_0-|z|^{-1}\bar{z}\partial_t\rho_2+2/3\partial_t\big[E_0(\rho_0^++\rho_0^-\big]\Big\}(\ldots)\,dz
  \\ \label{3-or1}
  -c^{-3}\int_{|z|\le ct}\int|z|^{-1}(\bar{z}\cdot p)\big(\Bar{z}\bar{z}\cdot p\big)\partial_t f_0(\ldots)\,dp\,dz 
\end{gather}
Using (\ref{rrVP}) and integration by parts  we compute  
\begin{gather}
  \label{3td3-or}
  1/3c^{-3}\int_{|z|\le ct} z\partial_t^3\rho_0(\ldots)\,dz
  \\ \nonumber
   = 1/3c^{-3}\int_{|z|\le ct} \partial_t^2j_0(\ldots)\,dz
     -1/3tc^{-2}\int_{|z|=ct} \bar{z}\bar{z}\cdot \partial_t^2 j_0(0, x+z)\,ds(z)+{\cal O}(c^{-4}).
\end{gather}
Summing the  terms with two time derivatives from (\ref{3-or1}) and (\ref{3td3-or}) we get
\begin{gather}
  \label{2td3-or}
  -2/3 c^{-3}\int_{|z|\le ct}\partial_t^2 j_0(\ldots)\,dz 
  = -2/3c^{-3}\int_{|z|\le ct}\partial_t\big[(E_0+(2/3)c^{-3}\Dd)(\rho_0^++\rho_0^-)\big](\ldots)\,dz
  \\ \nonumber
  +2/3c^{-3}\int_{|z|=ct}\int\bar{z}\cdot p \,p\,\partial_t f_0(0, x+z, p)\,dp\,ds(z)+{\cal O}(c^{-4}).
\end{gather}
Now we collect the first time derivatives from (\ref{3-or1}). 
Starting with the contributions containing $f_0$ we have
\begin{eqnarray}
  \lefteqn{-c^{-3}\int_{|z|\le ct}\int|z|^{-1}\big(\bar{z}(\bar{z}\cdot p)^2
  -(\bar{z}\cdot p) p\big)\partial_t f_0(\ldots, p)\,dp\,dz} 
  \nonumber \\
  & = & -c^{-3}\int_{|z|\le ct}\int|z|^{-1}\big(2(\bar{z}\cdot p)\bar{z}\otimes\bar{z}-\bar{z}\cdot p -p
  \otimes\bar{z}\big)
  \nonumber \\
   & &
  \hspace{10em}\Big(E_0(\ldots)+(2/3)c^{-3}\Dd(\hat{t}(z))\Big)(f^+_0+f^-_0)(\ldots,
  p)\,dp\,dz
  \nonumber \\
   & & -c^{-3}\int_{|z|\le
  ct}\int|z|^{-2}f_0(\ldots, p)\big(-4\bar{z}(\bar{z}\cdot p)^3+3p(\bar{z}\cdot p)^2
  +2\bar{z}(\bar{z}\cdot p)p^2-pp^2\big)\,dp\,dz
  \nonumber \\
  & & +c^{-3}(ct)^{-1}\int_{|z|=ct}\int\big(\bar{z}(\bar{z}\cdot p)^3-p(\bar{z}\cdot p)^2\big)\fon(x+z,
  p)\,dp\,ds(z) + {\cal O}(c^{-4})
  \label{1td3-or1}
\end{eqnarray}
and secondly, for the terms containing $f_2$ we compute employing (\ref{LVP})
\begin{eqnarray}
   \lefteqn{-c^{-3}\int_{|z|\le ct}|z|^{-1}\bar{z}\partial_t\rho_2(\ldots)\,dz}
   \nonumber\\
   & = & -c^{-3}\int_{|z|\le
   ct}\int|z|^{-2}f_2(\ldots, p)\big(-
   2\bar{z}(\bar{z}\cdot p)+p\big)\,dp\,dz
   \nonumber \\ 
   & & +c^{-3}(ct)^{-1}\int_{|z|=ct} \int
   \bar{z}(\bar{z}\cdot p) f^{\circ}_2(x+z, p)
   \,dp\,ds(z)
   \nonumber \\
   & & +1/2c^{-3}\int_{|z|\le
   ct}\int|z|^{-2}f_0(\ldots, p)\big(-2\bar{z}(\bar{z}\cdot p)p^2+p^2 p\big)\,dp\,dz
   \nonumber \\ 
    & & -1/2 c^{-3}(ct)^{-1}\int_{|z|=ct}\int\bar{z}(\bar{z}\cdot p)p^2\fon(x+z, p)\,dp\,ds(z)
   +{\cal O}(c^{-4}).
   \label{1td3-or2}
\end{eqnarray}
Summarizing (\ref{3-or1})-(\ref{1td3-or2}) we can identify the contribution in the third order of $\widetilde{E^R_{int}}$.
\begin{gather}
  \int_{|z|=ct}\int\big\{-1/3 tc^{-2}\bar{z}(\bar{z}\cdot p)\partial_t^2+2/3 c^{-3}(\bar{z}\cdot
  p)p\partial_t
  \nonumber \\
  +c^{-3}(ct)^{-1}\Big(\bar{z}(\bar{z}\cdot p)^3-p(\bar{z}\cdot p)^2-1/2\bar{z}(\bar{z}\cdot
  p)p^2\Big)\big\}f_0(0, x+z, p)
  \,dp\,ds(z)
  \nonumber \\
  +c^{-3}(ct)^{-1}\int_{|z|=ct} \int
   \bar{z}(\bar{z}\cdot p) f^{\circ}_2(x+z, p)
   \,dp\,ds(z)
  \nonumber \\
  +c^{-3}\int_{|z|\le ct}\int|z|^{-2}f_0(\ldots, p)\Big\{4\bar{z}(\bar{z}\cdot p)^3-3p(\bar{z}\cdot
  p)^2-3\bar{z}(\bar{z}\cdot p)p^2+3/2p p^2\Big\}\,dp\,dz
  \nonumber \\
  -c^{-3}\int_{|z|\le ct}\int|z|^{-1}\big\{2(\bar{z}\cdot p)\bar{z}\otimes\bar{z}-\bar{z}\cdot p -p
  \otimes\bar{z}\big\}E_0(\ldots)(f^+_0
  +f^-_0)(\ldots, p )\,dp\,dz
  \nonumber \\
  +c^{-3}\int_{|z|\le ct}\int|z|^{-2}f_2(\ldots, p)\big\{2\bar{z}(\bar{z}\cdot p)-p\big\}\,dp\,dz. 
  \label{3-or}
\end{gather}
\end{subequations}
Therefore, if we use (\ref{0-or}), (\ref{1-or}), (\ref{2-or}), (\ref{3-or}) and add some terms of order ${\cal
  O}(c^{-4})$ containing $f^\pm_2$ as e.g. $\int_{|z|\le ct}\int|z|^{-2}\bar{z}(\bar{z}\cdot p)^2 f_2 (\ldots)\,dp\,dz$ 
 it turns out that $E^R$ can be decomposed
as it is claimed in (\ref{ER-rep}). 

Similar calculations for $B^R$ yield 
\begin{equation}
  \label{BR-rep}
    B^R  = B^R_{ext}+B^R_{int}+B^R_{bound}+{\cal O}_{cpt}(c^{-4})
\end{equation}
where
\begin{eqnarray*}
  \label{BRext-rep}
  B^R_{ext}(t, x) & = &  c^{-1}\int_{|z|\ge ct}\Big(|z|^{-2}\bar{z}\times (j_0+ c^{-2} j_2)-1/2  c^{-2}\bar{z}\partial_t^2\jn\big)\txz\dz
  \\ \label{BRint-rep}
  B^R_{int}(t, x) & = & \int_{|z|\le ct}\int|z|^{-2} L_1(\bar{z}, c^{-1} p)f_R(\tdz, x+z, p)\diffp\dz
  \\\nonumber 
  & & + c^{-2}\int_{|z|\le ct}\int|z|^{-1} L_2(\bar{z}, c^{-1} p)\En(f^+_R+f^-_R)(\tdz, x+z, p)\diffp\dz
  \\ \label{BRbound-rep}
  B^R_{bound}(t, x) & = &\ctme\int_{|z|=ct}\int L_3(\bar{z}, c^{-1} p)\fon\xzp\diffp\dsz
\end{eqnarray*}
and the kernels are given by
\begin{eqnarray*}
  \label{L1}
  L_1(\bar{z},\tilde{p}) & = & \bar{z}\times\tilde{p}-2\bar{z}\times\tilde{p}(\bar{z}\cdot\tilde{p})3/2\bar{z}\times\tilde{p}\tilde{p}^2+3\bar{z}\times\tilde{p}(\bar{z}\cdot\tilde{p})^2
  \\ \label{L2}
  L_2(\bar{z},\tilde{p}) & = &\bar{z}\times(\cdots)-\bar{z}\cdot\tilde{p}\bar{z}\times(\cdots)-\bar{z}\cdot\tilde{p}\bar{z}\cdot(\cdots)\in \R^{3\times3}
  \\ \label{L3}
  L_3(\bar{z}, \tilde{p}) & = &  -\bar{z}\times\tilde{p}\bar{z}\cdot\tilde{p}+\bar{z}\times\tilde{p}(\bar{z}\cdot\tilde{p})^2.
\end{eqnarray*}
\subsection{Representation of the Maxwell fields $E$ and $B$}\label{repmax-sec}
In this section we will verify the representation formula (\ref{E-rep})
for the full Maxwell field $E$ by expanding the respective expressions
from \cite{glstr,schaeffer:86} to higher orders. Once again
the computation for the corresponding magnetic field $B$ is very similar
and therefore omitted. Let $(f, E, B)$ be a $C^1$-solution
of (\ref{RVMC}) with initial data $(f^{\circ, \pm}_c, E^\circ_c, B^\circ_c)$ according to (\ref{IC}). We recall the following
representation from \cite[(A13), (A14), (A3)]{schaeffer:86},
\begin{equation}
   E =  E_D+E_{DT}+E_T+E_S, \label{EFeld} 
\end{equation}
where
\begin{eqnarray*}
   E_D(t, x) & = &
   \partial_t\bigg(\frac{t}{4\pi}\int_{|\omega|=1} E^\circ_c(x+ct\omega)\,d\omega\bigg)
   +\frac{t}{4\pi}\int_{|\omega|=1}\partial_t E(0, x+ct\omega)\,d\omega, \\
   E_{DT}(t, x) & = & (ct)^{-1}\int_{|z|=ct}\int K_{DT}(\bar{z}, \hat{p})
   \fon(x+z, p)\,dp\,ds(z), \\
   E_T(t, x) & = & \int_{|z|\leq ct} |z|^{-2}\int K_T(\bar{z}, \hat{p})
   f(\ldots, p)\,dp\,dz, \\
   E_S(t, x) & = &c^{-2}\int_{|z|\leq ct} |z|^{-1}\int K_S(\bar{z}, p)
   (E+c^{-1}\hat{p}\times
   B)(\ldots)(f^++f^-)(\ldots, p)\,dp\,dz
\end{eqnarray*} 
The kernels are given by
\begin{subequations}
\begin{eqnarray}\nonumber
   K_{DT}(\bar{z}, \hat{p}) & = & -(1+c^{-1}\bar{z}\cdot\hat{p})^{-1}
   (\bar{z}-c^{-2}(\bar{z}\cdot\hat{p})\hat{p}), 
   \\ \label{KT-def}
   K_T(\bar{z}, \hat{p})  & = & -(1+c^{-1}\bar{z}\cdot\hat{p})^{-2}(1-c^{-2}\hat{p}^2)
   (\bar{z}+c^{-1}\hat{p}), 
   \\ \nonumber 
   K_S(\bar{z}, p)  & = & -(1+c^{-1}\bar{z}\cdot\hat{p})^{-2}(1+c^{-2} p^2)^{-1/2}
   \Big[(1+c^{-1}\bar{z}\cdot\hat{p})+c^{-2}((\bar{z}\cdot\hat{p})\bar{z}-\hat{p})\otimes\hat{p}
   \\ & & \hspace{14em} -(\bar{z}+c^{-1}\hat{p})\otimes\bar{z}\Big]\in\R^{3\times 3}
   \label{KS-def}
\end{eqnarray}
\end{subequations}
Next we expand these kernels in powers of $c^{-1}$. According to (\ref{SchrankeSupport}) the
$p$-support of $f^\pm(t, x, \cdot)$ is uniformly bounded in $x\in\R^3$
and $t\in [0, T]$  by $M_5$. Thus, we may suppose that $|p|\le M_5(T)$ in each of the $p$-integrals. Therefore as long
as $f^\pm(\ldots, p)\not=0$ we have
\begin{eqnarray*}
  K_T(\bar{z}, \hat{p}) & = & K_1(\bar{z}, c^{-1}p)+{\cal O}(c^{-4}) \\
  K_S(\bar{z}, \hat{p}) & = & K_2(\bar{z}, c^{-1}p)+{\cal O}(c^{-2}) \\
  K_{DT}(\bar{z}, \hat{p}) & = & -\bar{z}+K_3(\bar{z}, c^{-1}p)+{\cal O}(c^{-4}).
\end{eqnarray*}
Recall that $f^{\circ, \pm}_c=0$ if $|x|\ge r_0$, see (\ref{dat-def}) and (\ref{IC});
utilizing this as well as 
\begin{eqnarray*}
   \lefteqn{-(ct)^{-1}\int_{|z|=ct}\int_{|p|\le p^\dagger} {\cal O}(c^{-4})
   {\bf 1}_{B(0, r_0)}(x+z)\,dp\,ds(z)}
   \\
   & = & \bigg(ct\int_{|\omega|=1}{\bf 1}_{B(0, r_0)}(x+ct\omega)\,ds(\omega)\bigg){\cal O}(c^{-4})
    =  {\cal O}(c^{-4})
\end{eqnarray*}
by \cite[Lemma 1]{schaeffer:86}, uniformly in $x\in\R^3$, $t\in [0, T]$, and $c\ge 1$, we arrive at
\begin{equation}
  \label{EDT-exp}
  E_{DT}(t, x) =
  (ct)^{-1}\int_{|z|=ct}\int(-\bar{z}+K_3(\bar{z},
  c^{-1}p))(\fon +c^{-2} f^\circ_2)(x+z, p)\,dp\,ds(z)+{\cal O}(c^{-4}).
\end{equation}
Concerning $E_T$, we note that $f^\pm(t, x, p)=0$
for $|x|\ge M_5(T)$ and $t\le T$, see (\ref{SchrankeSupport}).
Since, by distinguishing the cases $|x-y|\ge 1$ and $|x-y|\le 1$,
\[ \int_{|z|\leq ct} |z|^{-2}\,{\bf 1}_{B(0, M_5(T))}(x+z)\,dz
   =\int_{|x-y|\leq ct} |x-y|^{-2}\,{\bf 1}_{B(0, M_5(T))}(y)\,dy={\cal O}(1) \]
uniformly in $x\in\R^3$, $t\in [0, T]$, and $c\ge 1$, similar computations
as before show that
\begin{subequations}
\begin{eqnarray}\label{ET-expa}
   E_T(t, x) & = & \int_{|z|\leq ct} |z|^{-2}\int
   K_1(\bar{z}, c^{-1}p)f(\ldots, p)\,dp\,dz+{\cal O}(c^{-4}).
\end{eqnarray}
In the same manner, elementary calculations using also (\ref{SchrankeFelder})
can be carried out to get
\begin{equation}
   E_S(t, x)  =  c^{-2}\int_{|z|\leq ct} |z|^{-1}\int K_2(\bar{z}, c^{-1}p)
   E(\ldots)(f^+ +f^-)(\ldots, p) \,dp\,dz +{\cal
   O}(c^{-4}).
   \label{ES-expa} 
\end{equation}
Observe that we have proved
\begin{equation}
  \label{maxEin-est}
  E_T+E_S=E_{int}+{\cal O}(c^{-4}),
\end{equation}
see (\ref{Eint-rep}).
\end{subequations}
Next we consider the data term
\begin{equation}\label{dat1} 
   E_D(t, x) = 
   \partial_t\big(\frac{t}{4\pi}\int_{|\omega|=1} E^\circ_c(x+ct\omega)\,d\omega\big)
   +\frac{t}{4\pi}\int_{|\omega|=1}\partial_t E(0,x+ct\omega)\,d\omega = I+II 
\end{equation}
By Maxwell's equations we have $\partial_t E(0, x)=c\nabla\times B^\circ_c (x)-4\pi j(0, x)$.
Recall that $B^\circ_c =c^{-1}B_1+c^{-3}B_3+c^{-4}B_{c, free}$.
Using
$\Delta=-\nabla\times\nabla\times+\nabla\nabla\cdot$
and $\partial_t\rho_0+\nabla\cdot j_0=0$ as well
as $\partial_t\rho_2+\nabla\cdot j_2=0$, the
latter an easy 
consequence of (\ref{LVP}), it is easy to check that 
$\nabla\times B_1=\partial_t E_0+4\pi j_0$ and $\nabla\times B_3=\partial_t E_2+4\pi j_2$. 
Employing the definition of $E_0$, $E_2$ and (\ref{formeli})-(\ref{formeliv}) below we compute
\begin{subequations}
\begin{eqnarray}
  \label{dte0}
  \lefteqn{\frac{t}{4\pi}\int_{|\omega|=1}\partial_t E_0(0, x+ct\omega)\,d\omega =
  -\frac{t}{4\pi}\int_{|\omega|=1}\int|z|^{-2}\bar{z}\partial_t\rho_0(0, x+ct\omega+z)\,dz\,d\omega}
  \nonumber \\
  & = & -\frac{t}{4\pi}\int\partial_t\rho_0(0, y)\int_{|\omega|=1}|y-x-ct\omega|^{-3}(y-x-ct\omega)\,d\omega\,dy
  \nonumber \\
  & = & -t\int_{|z|\ge ct}|z|^{-2}\bar{z}\partial_t\rho_0(0, x+z)\,dz
  \\ \label{dte2}
  \lefteqn{\frac{t}{c^24\pi}\int_{|\omega|=1}\partial_t E_2(0, x+ct\omega)\,d\omega} 
  \\ \nonumber 
  & = &
  \frac{t}{c^24\pi}\int_{|\omega|=1}\int\big\{1/2\bar{z}\partial_t^3\rho_0-|z|^{-1}\partial_t^2 j_0-|z|^{-2}\bar{z}\partial_t\rho_2\big\}(0,
  x+ct\omega+z)\,dz\,d\omega
  \nonumber \\
  & = & \frac{1}{3c^3}\int_{|z|\le ct} z\partial_t^3\rho_0(0, x+z)\,dz+\frac{t}{2c^2}\int_{|z|\ge ct} \bar{z}\partial_t^3\rho_0(0,
  x+z)\,dz
  \nonumber \\
  & &-\frac{t^3}{6}\int_{|z|\ge ct}|z|^{-2}\bar{z}\partial_t^3\rho_0(0, x+z)\,dz
  -\frac{1}{c^3}\int_{|z|\le ct}\partial_t^2 j_0(0, x+z)\,dz
  \nonumber \\
  & & -\frac{t}{c^2}\int_{|z|\ge ct}|z|^{-1}\partial_t^2 j_0(0, x+z)\,dz-
  \frac{t}{c^2}\int_{|z|\ge ct}|z|^{-2}\bar{z}\partial_t\rho_2(0, x+z)\,dz.
  \nonumber
\end{eqnarray}
Using (\ref{rrVP}) we also have
\begin{eqnarray}
  \label{dte2-1}
  \lefteqn{\frac{1}{3c^3}\int_{|z|\le ct} z\partial_t^3\rho_0(0, x+z)\,dz-\frac{1}{c^3}\int_{|z|\le ct}\partial_t^2 j_0(0, x+z)\,dz}
  \\
  & = & -\frac{t}{3c^2}\int_{|z|=ct}\bar{z}\bar{z}\cdot \partial_t^2 j_0(0, x+z)\,dz-\frac{2}{3c^3}\int_{|z|\le ct}\partial_t^2 j_0(0, x+z)\,dz
  \nonumber \\
  & = & -\frac{t}{3c^2}\int_{|z|=ct}\bar{z}\bar{z}\cdot \partial_t^2 j_0(0, x+z)\,dz+\frac{2}{3c^3}\int_{|z|=ct}\int(\bar{z}\cdot p)\partial_t j_0(0,
  x+z)\,dz
  \nonumber \\
  &&-\frac{2}{3c^3}\int_{|z|\le ct}\partial_t\big[(E_0+(2/3)c^{-3}\Dd)(\rho^+_0+\rho^-_0)\big](0, x+z)\,dz.
  \nonumber 
\end{eqnarray}
Furthermore, according to (\ref{dat-def}) and
(\ref{freedat-def}) and the definition of $j_0$
and $j_2$ the support of
$\tilde{j}:=(j-j_0-c^{-2}j_2)(0, \cdot)$ is
bounded by $r_0$ and  
\[ \tilde{j}(x)=\int_{|p|\le r_0}
\Big[(\hat{p}-p+\frac{p^2}{2c^2})\fon+
  c^{-2}(\hat{p}-p)f^\circ_2
  +c^{-4}\hat{p}f^\circ_{c, free} \Big](x, p)\,dp = {\cal
  O}(c^{-4}).\]
Utilizing \cite[Lemma 1]{schaeffer:86} we conclude
\begin{equation}
  \label{jdat}
\frac{ct}{4\pi}\int_{|\omega|=1}4\pi\tilde{j}(0, x+ct\omega)\,d\omega = {\cal O}(c^{-4}).
\end{equation}
\end{subequations}
Now we return to the contributions coming from I in (\ref{dat1}).
Note that in \cite[p304/305,(5.21)]{bauku} the contributions  coming from $E_0$ and $E_2$ are already determined to
\begin{subequations}
\begin{gather}
  -\int_{|z|>ct} |z|^{-2}\bar{z}\,\rho_0(0, x+z)\,dz
  +\frac{1}{2}\,c^{-2}\int_{|z|>ct}\bar{z}\,\partial_t^2\rho_0(0,
  x+z)\,dz
  \nonumber \\  
  -\frac{1}{2}\,t^2\int_{|z|>ct}|z|^{-2}\bar{z}
  \,\partial_t^2\rho_0(0, x+z)\,dz
  -c^{-2}\int_{|z|>ct}|z|^{-1}\partial_t j_0(0, x+z)\,dz
  \nonumber \\ 
  +(ct)^{-1}\int_{|z|=ct} \bar{z}\,(\rho_0
  +c^{-2}\rho_2)(0, x+z)\,ds(z).
  \label{Idat}
\end{gather}
Concerning the two remaining terms we have
\begin{eqnarray}
  \label{D3dat}
  \lefteqn{\partial_t\bigg(\frac{t}{4\pi}\int_{|\omega|=1}\frac{2}{3c^3}D^{[3]}(0)\,d\omega\bigg) = \frac{2}{3c^3}D^{[3]}(0)
  =\frac{2}{3c^3}\int\partial_t\big[E_0(\rho_0^++\rho_0^-)\big](0, z)\,dz}
  \\ \nonumber 
  & = & 
   \frac{2}{3c^3}\int_{|z|\ge ct} \partial_t\big[E_0(\rho_0^++\rho_0^-)\big](0, x+z)\,dz 
   +\frac{2}{3c^3}\int_{|z|\le ct} \partial_t\big[E_0(\rho_0^++\rho_0^-)\big](0, x+z)\,dz,
\end{eqnarray}
note that the last term in (\ref{D3dat}) cancels the last term in (\ref{dte2-1});
and 
\begin{eqnarray}
  \label{efree-dat}
  \lefteqn{\partial_t\big(\frac{t}{4\pi}\int_{|\omega|=1}c^{-4}E^\circ_{c, free}(x+ct\omega)\,d\omega\big)}
  \\ \nonumber
  & = & \frac{c^{-4}}{4\pi}\int_{|\omega|=1}E^\circ_{c, free}(x+ct\omega)\,d\omega+\frac{ct}{c^4}\int_{|\omega|=1}DE_{c,
  free}(x+ct\omega)\omega\,d\omega
  = {\cal O}(c^{-4})
\end{eqnarray}
according to (\ref{freedat-def}), \cite[Lemma
1]{schaeffer:86}. 
\end{subequations}                                                            
Thus, combining (\ref{EDT-exp}) and (\ref{dat1})-(\ref{efree-dat}) the representation formulas (\ref{Eext-rep}) and
(\ref{Ebound-rep}) are proved.
\subsubsection{Representation of $B$}\label{B-ref-sec}
We simply state the representation of $B$.
\begin{equation}
  \label{B-rep}
  B = B_{ext}+B_{int}+B_{bound}+{\cal O}(c^{-4})
\end{equation}
\begin{eqnarray}
  \nonumber
  B_{ext}(t, x) & = &
  c^{-1}\int_{|z|\le ct}\big\{|z|^{-2}\bar{z}\times\big(j^R-t\partial_t\jn-t^2/2\jn\big)
   \\ \nonumber 
   & & -1/2 c^{-2}\bar{z}\times\partial_t^2\jn\big\}(0, x+z)\dz
  \\ 
  \nonumber
  B_{int}(t, x) & = & \int_{|z|\le ct}\int|z|^{-2} L_1(\bar{z}, c^{-1} p)f(\tdz, x+z, p)\diffp\dz
  \\ \nonumber 
  & & + c^{-2}\int_{|z|\le ct}\int|z|^{-1} L_2(\bar{z}, c^{-1} p) E (f^++f^-)(\tdz, x+z, p)\diffp\dz
  \\ 
  \nonumber
  B_{bound}(t, x) & = & B^R_{bound}
\end{eqnarray}

\subsection{Some explicit integrals and a lemma}

We point out some formulas that have been used in the previous sections.
For $z\in\R^3$ and $r>0$ an elementary calculation yields
\begin{subequations}\begin{equation}\label{formeli}
  \int_{|\omega|=1}|z-r\omega|^{-1}\,d\omega
  =\left\{\begin{array}{c@{\quad:\quad}c}
  4\pi r^{-1} & r\geq |z| \\ 4\pi |z|^{-1} & r\leq |z|
  \end{array}\right. .
\end{equation}
Differentiation w.r.t.~$z$ gives
\begin{equation}\label{formelii}
   \int_{|\omega|=1}|z-r\omega|^{-3}(z-r\omega)\,d\omega
   =\left\{\begin{array}{c@{\quad:\quad}c}
   0 & r>|z| \\ 4\pi |z|^{-2}\bar{z} & r<|z|
   \end{array}\right. .
\end{equation}
Similarly,
\[ \int_{|\omega|=1}|z-r\omega|\,d\omega
   =\left\{\begin{array}{c@{\quad:\quad}c}
   4\pi r+\frac{4\pi}{3}z^2 r^{-1} & r\geq |z| \\[1ex]
   4\pi |z|+\frac{4\pi}{3}r^2 |z|^{-1} & r\leq |z|
   \end{array}\right. , \]
and thus by differentiation
\begin{equation}\label{formeliv}
   \int_{|\omega|=1}|z-r\omega|^{-1}(z-r\omega)\,d\omega
   =\left\{\begin{array}{c@{\quad:\quad}c}
   \frac{8\pi}{3r}\,z & r>|z| \\[1ex] 4\pi\bar{z}
   -\frac{4\pi}{3}r^2|z|^{-2}\bar{z} & r<|z|\end{array}\right. .
\end{equation}
Finally, for $z\in\R^3\setminus\{0\}$ also
\begin{equation}\label{formelv}
  \int |z-v|^{-1}|v|^{-3}v\,dv=2\pi\bar{z}
\end{equation}
can be computed. 
\end{subequations}

\bigskip\bigskip

\noindent
{\bf Acknowledgments:} The author is indebted to M.~Kunze, G.~Rein, A.~Rendall
and H.~Spohn for many discussions.

\end{document}